\documentclass[aps,prd,reprint,groupedaddress,floatfix,showpacs]{revtex4-1}

\usepackage{amsmath}
\usepackage{graphicx}
\usepackage{dcolumn}
\usepackage{bm}
\usepackage{caption}
\usepackage{subcaption}

\begin{document}


\title{Testing General Relativity's No-Hair Theorem with X-Ray Observations of Black Holes}


\author{Janie K. Hoormann} %
\email{jhoormann@wustl.edu}
\author{Banafsheh Beheshtipour}
\author{Henric Krawczynski}
\affiliation{Washington University in St. Louis, Physics Department and McDonnell Center for the Space Sciences \\ One Brookings Drive, CB 1105, St. Louis, MO 63130}


\date{\today}

\begin{abstract}
Despite its success in the weak gravity regime, General Relativity (GR) has yet to be verified in the regime of strong gravity. In this paper, we present the results of detailed ray tracing simulations aiming at clarifying if the combined information from X-ray spectroscopy, timing and polarization observations of stellar mass and supermassive black holes can be used to test GR's no-hair theorem. The latter states that stationary astrophysical black holes are described by the Kerr-family of metrics with the black hole mass and spin being the only free parameters.  We use four ``non-Kerr metrics'', some phenomenological in nature and others motivated by alternative theories of gravity, and study the observational signatures of deviations from the Kerr metric. Particular attention is given to the case when all the metrics are set to give the same Innermost Stable Circular Orbit (ISCO) in quasi-Boyer Lindquist coordinates. We give a detailed discussion of similarities and differences of the observational signatures predicted for BHs in the Kerr metric and the non-Kerr metrics. We emphasize that even though some regions of the parameter space are nearly degenerate even when combining the information from all observational channels, X-ray observations of very rapidly spinning black holes can be used to exclude large regions of the parameter space of the alternative metrics. Although it proves difficult to distinguish between the Kerr and non-Kerr metrics for some portions of the parameter space, the observations of very rapidly spinning black holes like Cyg X-1 can be used to rule out large regions for several black hole metrics.
\end{abstract}


\maketitle

\section{Introduction \label{sec:intro}}
In the early 1900's, Albert Einstein proposed his now famous theory of General Relativity.  Since its introduction, General Relativity (GR) has been tested extensively in our solar system. GR has passed all tests with remarkable accuracy, including but not limited to the perihelion shift of Mercury, the deflection of light passing near the Sun, and the Shapiro time delay.  
These tests were extended beyond the solar system with the discovery of the Hulse-Taylor binary pulsar where the decay rate of the orbital period was found to be consistent with the expected decay due to the loss of energy via gravitational waves (see \cite{Will2006} for a review of the subject).  However, despite all of these successes, GR has yet to be verified in the strong gravity regime. The most extreme gravitational fields are found near black holes (BHs), and 
tests of GR near BHs have received considerable attention  (see \cite{Psaltis2008} and references therein).  Much of this work makes use of the no-hair theorem of GR which states that the only stationary axially symmetric solutions of the Einstein equations are given by the Kerr/Newman family of metrics. In the case of astrophysical BHs with negligible electrical 
charge, the Kerr solutions are parameterized by the BH's geometric mass, $M$, and spin, $a$.  
The tests of this theorem include using stars orbiting Sagittarius $A^{*}$, the supermassive BH in the center of the Milky Way galaxy, to measure its angular momentum and quadrupole moment  \cite{Will2008}.
Gravitational wave observations of merging black holes have the potential to test strong-field GR not only in the stationary, but also
in the dynamic regime \cite{Dreyer2004}.  The work presented in this paper makes use of several recently proposed metrics that contain additional terms which violate the no-hair theorem including those of \cite{Johannsen2011a, Glampedakis2006, Aliev2005,Pani2011}. These metrics are used to quantify the degree to which spectroscopic, polarimetric, and timing X-ray observations can constrain deviations from the Kerr metric.

The spectral X-ray emission from stellar mass BHs can be characterized by a thermal continuum emitted from the accretion disk along with a power law component originating from hotter - yet mostly thermal - plasma,  commonly referred to as the corona \cite[e.g.][]{Gilfanov2014}. 
The power law emission is modeled using the equation $N(E) \propto E^{-\Gamma}$ where $\Gamma$ is the photon index \cite{Remillard2006}.  This emission can either travel directly to the observer or return to the accretion disk and lead to
scattered and/or reprocessed emission.  The reflection component contains the Fe-K$\alpha$ line at 6.4 keV 
and the Compton hump at energies greater than 20 keV.  The prominent Fe-K$\alpha$ line comes from the reprocessing of photons in the inner accretion disk and receives its characteristic broadened profile due to gravitational redshift, Doppler effects, and relativistic beaming.  This profile can then be fit to determine the BH's spin \cite{Fabian2000,Brenneman2013,Reynolds2003}.  The thermal emission, modeled using the prescription developed by \cite{Novikov1973} for a geometrically thin, optically thick disk, can also be used to deduce the spin of stellar mass black holes because it allows one to determine the radius of the Innermost Stable Circular Orbit (ISCO) which is a monotonic function of its spin \cite{McClintock2011,McClintock2014}.  In the case of supermassive black holes, the thermal disk emission falls into the optical/UV bands. As many different emission components contribute to the observed emission in these bands, fitting the Fe-K$\alpha$ line is the only way to determine the spin of these systems.

Polarimetric observations with photoelectric effect polarimeters like the ones used on the IXPE, PRAXyS, and XIPE missions
currently studied by NASA (IXPE \cite{Weisskopf2014} and PRAXyS \cite{Jahoda2015}) and ESA (XIPE \cite{Soffitta2013}) or with scattering polarimeters such as {\it PolSTAR} \cite{Beilicke2012, Beilicke2014, Guo2013} provide a new way to study inner structure of accretion flows.
Polarimeters can provide geometrical information even though the inner accretion flow of most black holes are 
too small to be imaged with the current or even next-generation telescopes 
(the only exception being the supermassive black holes Sgr A$^*$ and M~87 which might be imaged with the 
Event Horizon Telescope \cite{Ricarte2015}).
In the case of stellar mass black holes, the thermal X-ray emission is expected to exhibit linear polarization 
with the polarization fraction being a function of the inclination of the inner accretion disk \cite{Li2009,Schnittman2009}.  
The polarization of the Comptonized emission from the corona depends on the scattering processes in the corona itself and off the accretion disk. Several corona geometries have been studied including the lamp-post model where photons are 
emitted from a point source directly above the black hole itself \cite{Dovciak2012}. 
Other corona models assume a wedge or spherical corona geometry surrounding the accretion disk \cite[e.g.\ ][and references therein]{Schnittman2010}. 

In the past several years X-ray reverberation has come into its own as a powerful tool to study accreting black holes. 
Corona emission scattering off the accretion disk reaches the observer with a time delay relative to the direct corona emission. The energy dependence of the time delays can be used to infer details about the structure of the inner accretion flow. For a sample of AGNs, Fe-K$\alpha$ vs. continuum lags have been established along with Compton hump vs. continuum lags (see \cite{Kara2015,Uttley2014} for a review of the subject). These lags can be fit with numerical models to deduce system parameters such as the inclination and lamp-post height in NGC 4151 \cite{Cackett2014} and an extended corona geometry in 1H0707-495 \cite{Wilkins2013}.

Several authors have used the alternative BH metrics to find observational signatures of non-GR effects.
The following types of observations have been studied:
(i) fitting of the thermal X-ray emission from stellar mass black holes \cite{Bambi2011,Bambi2012a,Pun2008};
(ii) fitting of the Fe-K$\alpha$ line emission from stellar mass and supermassive black holes \cite{Bambi2013a,Bambi2013b, Johannsen2013b,Psaltis2012}; 
(iii) spectropolarimetric  observations of stellar mass black holes \cite{Krawczynski2012, Liu2015};
(iv) observations of Quasi-Periodic Oscillations (QPOs) \cite{Johannsen2011b, Bambi2012b, Johannsen2014};
(v) X-ray reverberation observations \cite{Jiang2015};
(vi) observations of the radiatively inefficient accretion flow around Sgr $A^*$ \cite{Broderick2013}. 
The studies showed that it is very difficult to observationally distinguish between the Kerr space time and 
the non-Kerr BH space times as long as the BH spin and the parameter describing the deviation from the Kerr space time
are free parameters that both need to be derived from the observations. 

Several approaches have been discussed  to break the degeneracy between the BH spin and 
deviation parameter(s). In \cite{Bambi2012a,Bambi2012c}, for example, it is proposed that the BH spin can be measured
independently from the accretion disk properties based on 
measuring the jet power, although this method faces several difficulties in practice \cite{Narayan2012}.
Furthermore, the observed results depend on the physics of the accretion disk, the radiation transport around the
black hole, the physics of launching and accelerating the jet, and the physics of converting the mechanical and
electromagnetic jet energy into observable electromagnetic jet emission. 

This paper follows up on the work of \cite{Krawczynski2012}.  The thermal emission from a geometrically thin,
optically thick accretion disk is modeled self-consistently for the Kerr metric and the alternative metrics,
and observational signatures are derived with the help of a ray-tracing code that tracks photons from their origin
to the observer, enabling the modeling of repeated scatterings of the photons off the accretion disk. 
This paper adds to the previous work by (i) covering the Kerr metric, the metric of Johannsen and Psaltis (2011) \cite{Johannsen2011a} and
three additional metrics, (ii) by modeling not only the thermal disk emission but also the emission from a lamp-post corona 
and the reprocessing of the coronal emission by the accretion disk, and (iii) by considering many observational channels.
We analyze the multi-temperature continuum emission from the accretion disk, 
the energy spectra of the reflected emission (including the Fe K-$\alpha$ line and the Compton hump), 
the orbital periods of matter orbiting the black hole close to the ISCO, the time lags between the
Fe K-$\alpha$ emission and the direct corona emission, and the size and shape of the black hole shadows.

The rest of the paper is organized as follows.  Section 2 begins with a summary of the alternative spacetimes used in this paper and goes on to 
discuss the model for both the thermal and coronal emission.  
In Section 3 we compare the observational signatures of the Kerr and the non-Kerr metrics finding that the observational
differences are rather small given the uncertainties about the properties of astrophysical accretion disks.
We summarize the results in Sect.\ 4 and emphasize that even though the Kerr and non-Kerr metrics 
can produce similar observational signatures for some regions of the respective parameter spaces, 
we can use X-ray observations of black holes from the literature to rule out large regions of the parameter 
space of the non-Kerr metrics. 

Throughout this paper we assume $c=\hbar=G=1$; all distances are given 
in units of the gravitational radius, $r_g = GM/c^2$.

\section{Methodology}
	\subsection{Alternative Metrics}
	As a way to test the no-hair theorem of general relativity, several non-Kerr metrics have been introduced which contain additional parameters apart from the BH's mass and spin. In this paper we employ the use of four non-GR metrics including two phenomenological metrics \cite{Johannsen2011a,Glampedakis2006} and two which are solutions to alternative theories of gravity \cite{Aliev2005,Pani2011}. All metrics are variations of the Kerr metric in (quasi) Boyer Lindquist coordinates $x^\mu = (ct, r, \theta, \phi)$. 
	
	The phenomenological metric of Johannsen and Psaltis, 2011 \cite{Johannsen2011a} (JP) reads:
	\begin{equation}
	\begin{aligned} \label{jpmetric}
	ds^2   & = -[1+h(r,\theta)]\left(1-\frac{2Mr}{\Sigma}\right)dt^2-\frac{4aMr\sin^2\theta}{\Sigma} \\ 
	       & \times [1+h(r,\theta)]dtd\phi  +\frac{\Sigma[1+h(r,\theta)]}{\Delta + a^2\sin^2\theta h(r,\theta)}dr^2 \\ 
	       & + \Sigma d\theta ^2 + \bigg[ \sin ^2 \theta \left( r^2+a^2+\frac{2a^2Mr\sin^2\theta}{\Sigma}\right) \\ 
	       & +h(r,\theta) \frac{a^2(\Sigma+2Mr)\sin^4\theta}{\Sigma} \bigg] d\phi^2
	\end{aligned}
	\end{equation}
	with
	\begin{equation} \label{sigma}
	\Sigma \equiv r^2 + a^2 \cos^2 \theta\ \\
	\end{equation} 
	\begin{equation} \label{delta}
	\Delta \equiv r^2 -2Mr +a^2.
	\end{equation}
	The metric was derived by modifying the temporal and radial components of the Schwarzschild line element by
	a term $h(r,\theta)$. The metric {\it does not exhibit any pathologies outside the event horizon} 
	and can be used for slowly and rapidly spinning black holes. 
	Asymptotic flatness constrains the leading terms of the expansion of $h$ in powers of $r$ 
	and the lowest order correction reads:
	\begin{equation}
	h(r,\theta)= \epsilon_3 \frac{M^3 r}{\Sigma ^2}.
	\end{equation}
	In the limit as $\epsilon_3 \rightarrow 0$ this metric reduces to the Kerr solution in Boyer-Lindquist coordinates. 
	
	Glampedakis and Babak, 2006 \cite{Glampedakis2006} (GB) introduced a metric for 
	slowly spinning black holes ($a \lesssim 0.4$).  This quasi-Kerr metric in Boyer-Lindquist coordinates is
	\begin{equation} \label{gbmetric}
	g_{ab}= g_{ab}^K + \epsilon h_{ab}
	\end{equation}
	where $g_{ab}^K$ is the Kerr metric and $h_{ab}$ is given by
	\begin{mathletters}
	\begin{eqnarray}
	h^{tt} &=& \left(1-\frac{2M}{r}\right)^{-1}[(1-3\cos^2\theta)\mathcal{F}_1(r)] \\ \nonumber
	h^{rr} &=& \left(1-\frac{2M}{r}\right)[(1-3\cos^2\theta)\mathcal{F}_1(r)] \\ \nonumber
	h^{\theta\theta} &=& -\frac{1}{r^2}[(1-3\cos^2\theta)\mathcal{F}_2(r)] \\ \nonumber
	h^{\phi\phi} &=& -\frac{1}{r^2\sin^2 \theta}[(1-3\cos^2\theta)\mathcal{F}_2(r)] \\ \nonumber
	h^{t\phi} &=& 0 \nonumber
	\end{eqnarray} 
	\end{mathletters}
	where $\mathcal{F}_1(r)$ and $\mathcal{F}_2(r)$ are defined in Appendix A in \cite{Glampedakis2006}.  It is clear to see that equation \ref{gbmetric} reduces to the Kerr metric when $\epsilon \rightarrow 0$.  The details of the JP and GB space times are described in \cite{Johannsen2013a}.

	Another solution is presented by Aliev and G{\"u}mr{\"u}k{\c c}{\"u}o{\v g}lu, 2005 \cite{Aliev2005}  describing an axisymmetric, stationary metric for a rapidly rotating black hole which is on a 3-brane in the Randall-Sundrum braneworld. 
	The metric turns out to be identical to GR's Kerr-Newman metric of an electrically charged spinning black hole, the only
	difference that $\beta$ is not the electrical charge but a ``tidal charge''. 
	The metric (referred to as KN-metric in the following) is given by:	
	\begin{equation}
	\begin{aligned} \label{agmetric}
	ds^2 & = - \left( 1-\frac{2Mr-\beta}{\Sigma}\right) dt^2 -\frac{2a(2Mr-\beta)}{\Sigma} \\
	     &  \times \sin^2 \theta dt d\phi +  \frac{\Sigma}{\Delta}dr^2 +\Sigma d\theta^2 + \bigg( r^2+a^2 \\
	     &+\frac{2Mr-\beta}{\Sigma} a^2 \sin^2\theta \bigg) \sin^2\theta d\phi^2 
	\end{aligned}
	\end{equation}
	with $\Sigma$ being the same as equation \ref{sigma} and with $ \Delta = r^2+a^2-2Mr+\beta $.  
	
	Pani et al, 2011 \cite{Pani2011} (PMCC) give a family of solutions for slowly rotating BHs derived augmenting the Einstein-Hilbert action by quadratic and algebraic curvature invariants coupling to a single scalar field. The action is given by the expression:
	\begin{equation}
	\begin{aligned}
	S & =  \frac{1}{16\pi}\int \sqrt{-g}d^4x [R-2 \nabla_a\phi \nabla^a\phi-V(\phi) \\
	    & +f_1(\phi)R^2 +f_2(\phi)R_{ab}R^{ab}+f_3(\phi)R_{abcd}R^{abcd} \\
	    & + f_4(\phi)R_{abcd} \,  ^* R^{abcd}] +S_{mat}[\gamma(\phi)g_{\nu\mu}, \Psi_{mat}]
	\end{aligned}
	\end{equation}
	where
	\begin{equation}
	f_i(\phi)=\eta_i+\alpha_i \phi +\mathcal{O}(\phi^2) 
	\end{equation}
	for $i=1-4$, $S_{mat}$ is the matter action containing a generic non-minimal coupling, and $V(\phi)$ is the scalar self potential.  When $\alpha_3 =0$ the metric reduces to the one for Chern-Simons gravity and when $\alpha_4=0$ it becomes the Gauss-Bonnet solution.

	\subsection{Thermal Accretion Disk Emission and Photon Propagation}
	The code models the emission of photons and their propagation self-consistently for the Kerr and non-Kerr metrics \cite{Krawczynski2012}. The radial emission profile of the geometrically thin, optically thick accretion disk is calculated based on the general solution of Novikov and Thorne \cite{Novikov1973}, the relativistic extension of the Shakura-Sunyaev equations \cite{Shakura1973}.
	Writing the considered Kerr or non-Kerr metric in the form 
	\begin{equation}
	ds^2 = -e^{2\nu} dt^2 +e^{2\psi}(d\phi-\omega dt)^2 + e^{2\mu}dr^2 +dz^2,
	\end{equation}
	the conservation of rest mass, angular momentum, and energy give the following disk brightness $F(r)$ in
	the rest frame of the emitting plasma \cite{Page1974}:
	\begin{equation} \label{eq:flux}
	F(r)=\frac{\dot{M_0}}{4 \pi}e^{-(\nu+\psi+\mu)}f(r)
	\end{equation}
	with
	\begin{equation}
	f(r) \equiv \frac{-p^t_{,r}}{p_\phi} \int\limits_{r_{ISCO}}^{r}\frac{p_{\phi,r}}{p^t}dr.
	\end{equation}
	Here, $\dot{M_0}$ is the time averaged rate of accretion. The solution assumes a vanishing torque at the ISCO. We calculate the $r_{ISCO}$ by finding the location where the energy is a minimum by solving the equation $\frac{dE}{dr} = 0$ for planar, circular orbits.
		
	We model photons emitted between $r_{ISCO}$ and 100 $r_g$ that are tracked until they either fall into 
	black hole or reach a coordinate stationary observer at $r=$ 10,000 $r_g$.
	The photon trajectories are calculated by integrating the geodesic equations:
	\begin{equation}
	\frac{d^2 x^\mu}{d \lambda ^{\prime 2}}= -\Gamma^\mu_{ \;\;\sigma \nu} \frac{d x^\sigma}{d \lambda ^\prime} \frac{d x^\nu}{d \lambda ^ \prime}
	\end{equation}
	with a fourth order Runge Kutta method. The $\Gamma^\mu_{ \;\;\sigma \nu}$'s are the Christoffel symbols and $\lambda^\prime$ is the affine parameter. Tracking photons forward in time makes it possible to model trajectories with
	multiple scattering events and/or with the absorption and re-emission of photons in the accretion disk and/or in the corona. The initial polarization and the change in polarization upon scattering are calculated with the help of Chandrasekhar's results for optically thick atmospheres \cite{Chandrasekhar1960}.  
	The polarization vector, \textbf{f} is parallel transported according to the following equation
	\begin{equation}
	\frac{df^\mu}{d \lambda^\prime} = -\Gamma^\mu_{ \;\;\sigma \nu} f^\sigma \frac{dx^\nu}{d\lambda^\prime}.
	\end{equation}
	A more detailed description of the code can be found in \cite{Krawczynski2012}.
	\subsection{Lamp-Post Corona Model}
\begin{figure}
		 \includegraphics[width=0.42\textwidth]{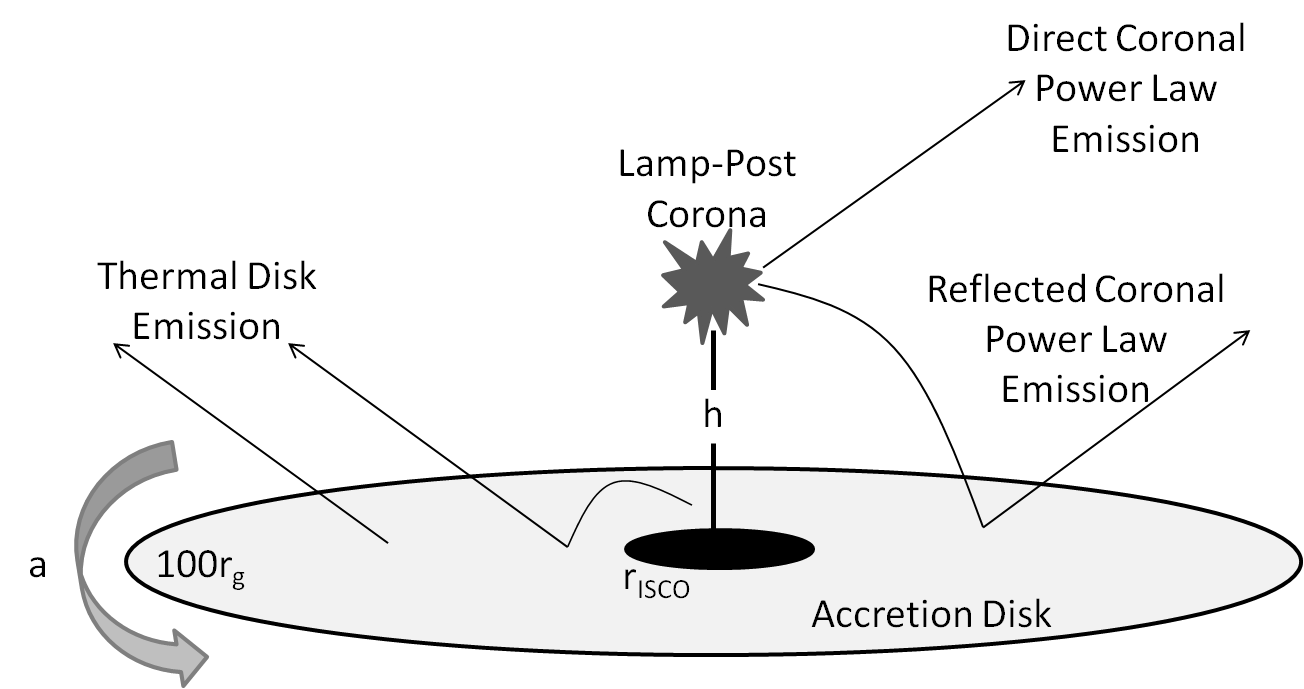}
		 \caption{\label{fig:diskDiagram} Diagram showing the thermal and coronal lamp-post emission surrounding a black hole.}
	\end{figure}

	We utilize  the commonly used lamp-post model \cite[e.g.][]{Matt1991,Dovciak2004} to simulate the coronal power-law emission. Unpolarized 1-100 keV photons are emitted from a point source above the black hole. 
		The height of the point source can be constrained based on fitting the Fe-K$\alpha$ vs. continuum emission lag \cite[e.g.][]{Cackett2014}. Alternatively, the corona may be associated with a region of energy dissipation close to the base of a jet. 
	The trajectory of the photons and their change in polarization upon scattering off the accretion disk are calculated as described above. Photons impinging on the accretion disk can either be absorbed, prompt the emission of a Fe-K$\alpha$ fluorescence photon, or Compton scatter.  A diagram of the various types of emission modeled can be seen in Fig. \ref{fig:diskDiagram}. 
	The results of \cite{George1991} were used to determine the probability of the creation of the Fe-K$\alpha$ photons and Compton reflections. 
	The results can then be analyzed to examine the time lag between the direct power-law emission and the Fe-K$\alpha$ emission from 2-10 keV following the methodology described in \cite{Cackett2014}. We analyze the Fourier transform of the transfer function to infer the time lag as a function of Fourier frequency.
	
\section{Results}
\begin{figure}
		       \centering
		       \begin{subfigure}[b]{0.43\textwidth}
		                \includegraphics[width=\textwidth]{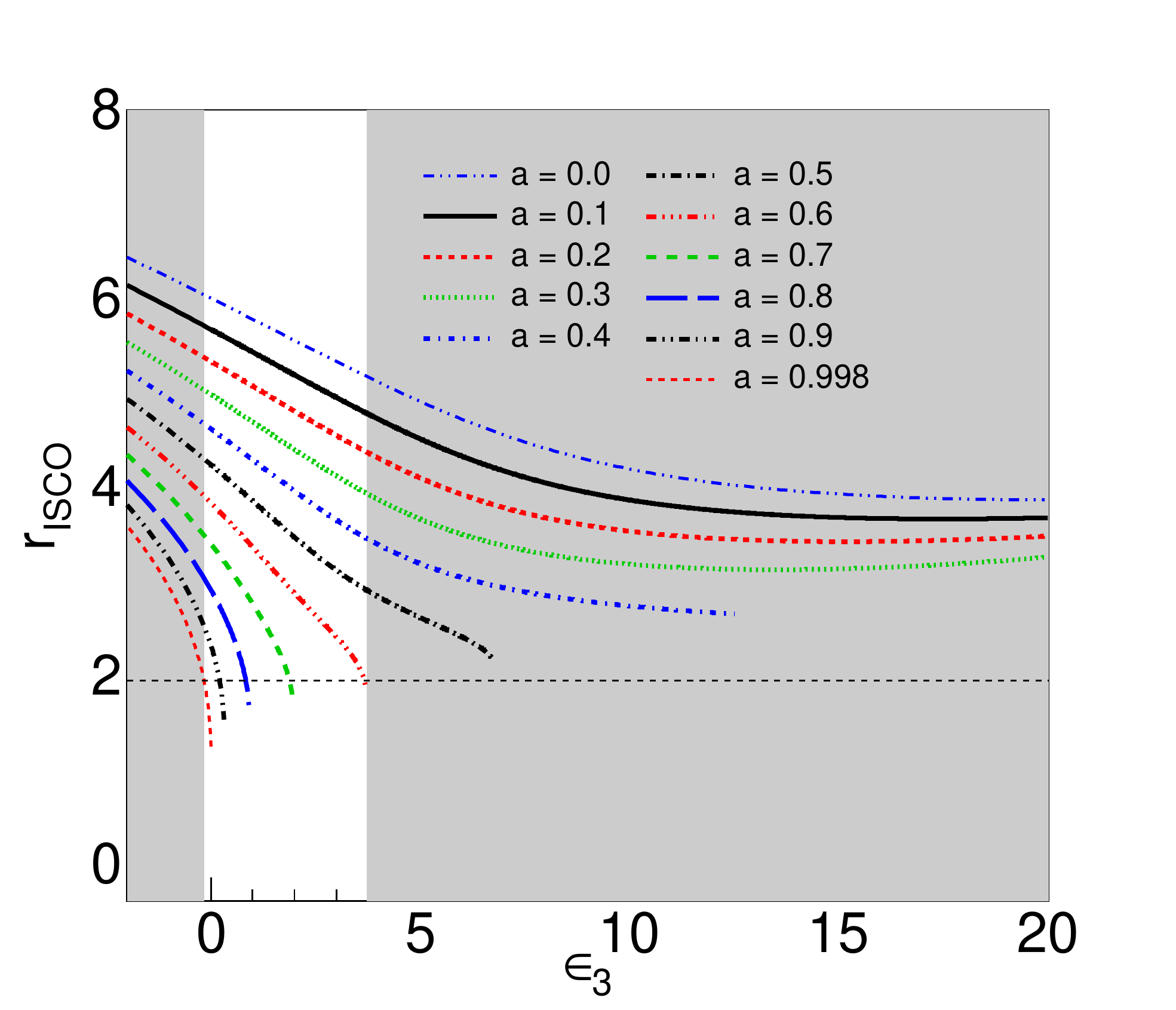}
		                \caption{\label{fig:familyJP}}
		        \end{subfigure}
		        \begin{subfigure}[b]{0.43\textwidth}
		                \includegraphics[width=\textwidth]{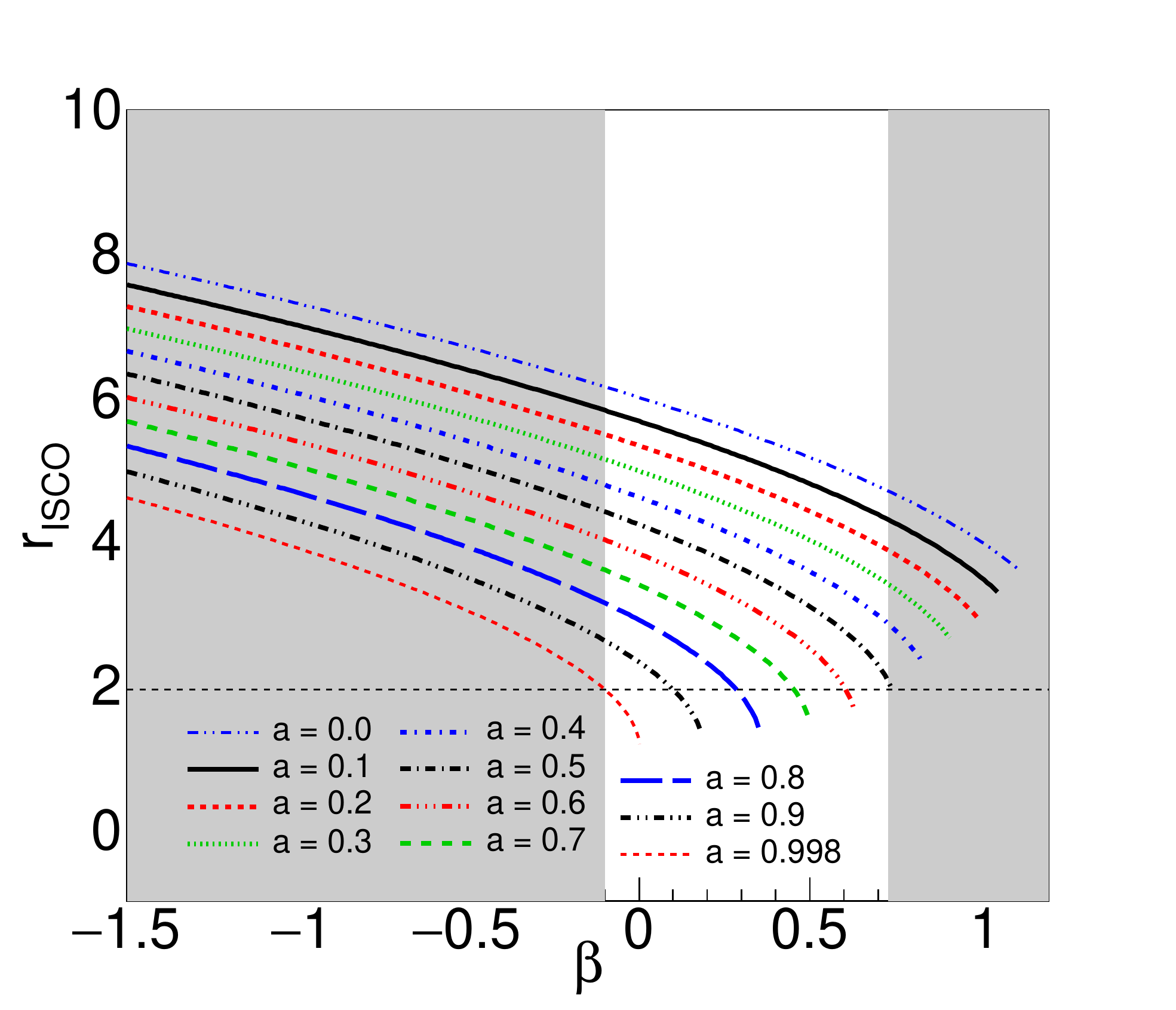}
		                \caption{\label{fig:familyKN}} 
		        \end{subfigure}
		       \caption{\label{fig:family}$r_{ISCO}$ as a function of deviation parameter for the JP metric (a) and the KN metric (b) with different spins illustrating the degeneracies within the metrics. The regions shaded grey indicate the portion of the parameter space excluded by observations of Cyg X-1 where the observed $r_{ISCO}$ \cite{Gou2011} is represented by the horizontal dashed black line (color online).}
			\end{figure}

Previous analyses have shown that the Kerr and non-Kerr metrics give similar spectral and spectropolarimetric 
signatures if the parameters are chosen to give the same $r_{ISCO}$ \cite[e.g.][]{Krawczynski2012,Bambi2013a,Jiang2015,Johannsen2014,Johannsen2013b,Kong2014}.  
Figure \ref{fig:family} 
shows $r_{ISCO}$ as a function of the BH spin $a$ and the parameter	
characterizing the deviation from
the Kerr metric for the JP and KN metrics.  We see that for all JP and KN metrics, we can always find one and only one 
Kerr metric with the same $r_{\rm ISCO}$. The mapping is not unambiguous the other way around: the JP and KN metrics 
can give one $r_{\rm ISCO}$ for several different combinations of the BH spin $a$ and the deviation parameter. 
	
In the following we focus on comparing ``degenerate'' models which give the same $r_{ISCO}$. 
We consider a slowly spinning Kerr black hole ($a=0.2, r_{ISCO}=5.33 r_g$) and a rapidly spinning Kerr black hole ($a=0.9, r_{ISCO}=2.32 r_g$) and JP and KN models giving the same $r_{ISCO}$ (see Table \ref{fig:table}).
\begin{table}
\caption{\label{fig:table} List of metric parameters used in the simulations.}
\begin{ruledtabular}
\begin{tabular}{lcccc}
\textrm{Metric}&
\textrm{Spin}&
\textrm{Deviation}&
\textrm{$r_{ISCO} $}&
\textrm{$\dot{M}$ (g/s)}\\
\colrule
Kerr & 0.9 & none & 2.32 & 8.98$\times 10^{17}$ \\ 
JP & 0.5 & $\epsilon_3 =$ 6.33 & 2.32 & 7.51$\times 10^{17}$\\
KN &0.5 & $\beta=$0.69 & 2.32 & 9.86$\times 10^{17}$\\ \hline
Kerr & 0.2 & none & 5.33 & 2.16$\times 10^{18}$ \\
GB & 0.25 & $\epsilon=0.12$ & 5.33 & 2.15$\times 10^{18}$\\
PMCC & 0.29 & $\alpha_3=0$, $\alpha_4=2.07$ & 5.33 & 2.11$\times 10^{18}$\\
\end{tabular}
\end{ruledtabular}
\end{table}
In the following we show the Kerr, GB and PMCC results for the low-spin case, 
and the Kerr, KN and JP results for the high-spin case. We adjusted the accretion rates to give the same 
accretion luminosity (extracted gravitational energy per unit observer time) for all considered metrics.  This is done by normalizing the accretion rate by the efficiency which is not corrected for the fraction of photons escaping to infinity.
	\begin{figure}
        \centering
        \begin{subfigure}[b]{0.21\textwidth}
                \includegraphics[trim=35mm 0mm 30mm 0mm ,width=\textwidth]{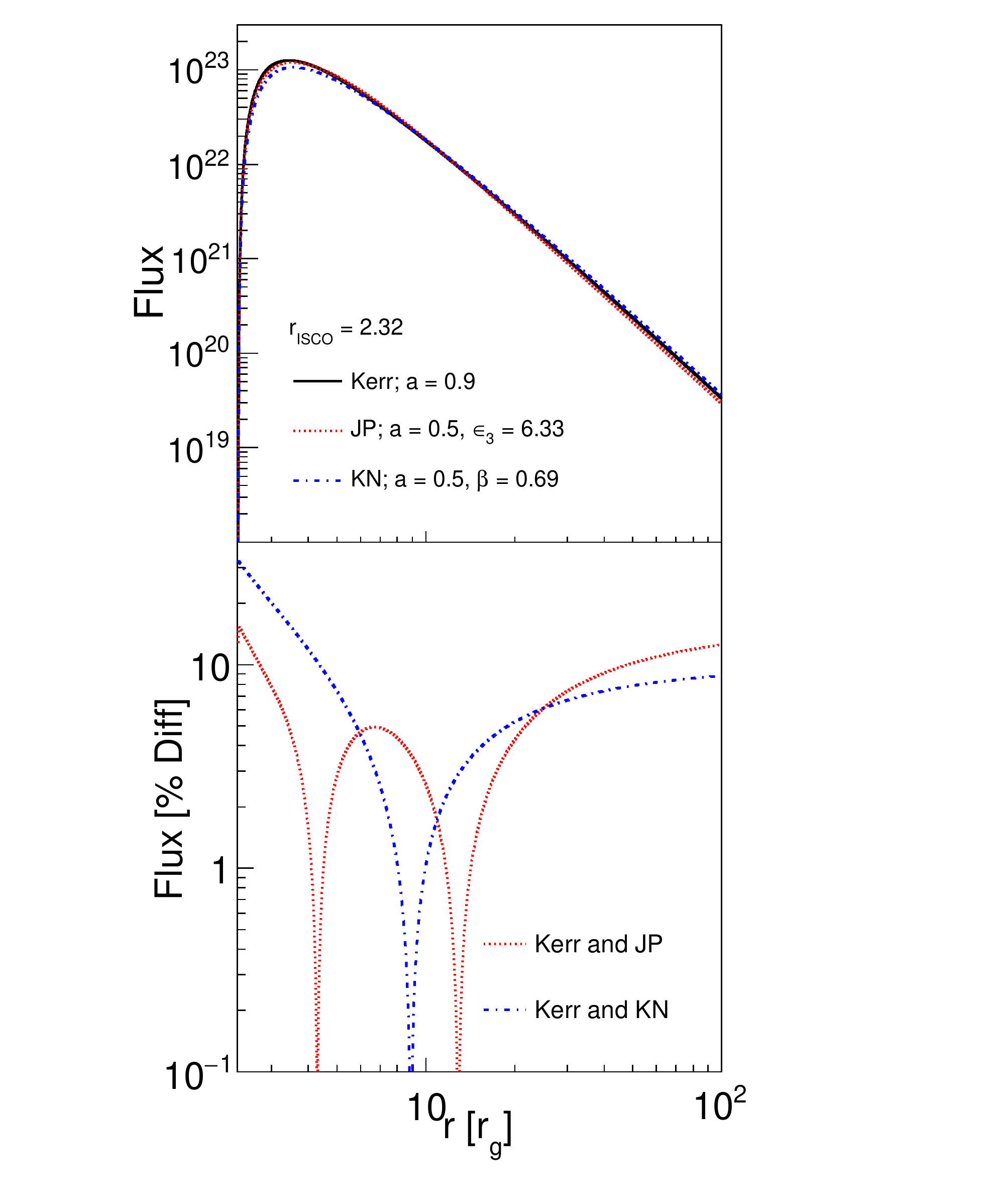}
                \caption{\label{fig:flux}}
        \end{subfigure}
         \quad
        \begin{subfigure}[b]{0.21\textwidth}
                \includegraphics[trim=30mm 0mm 35mm 0mm ,width=\textwidth]{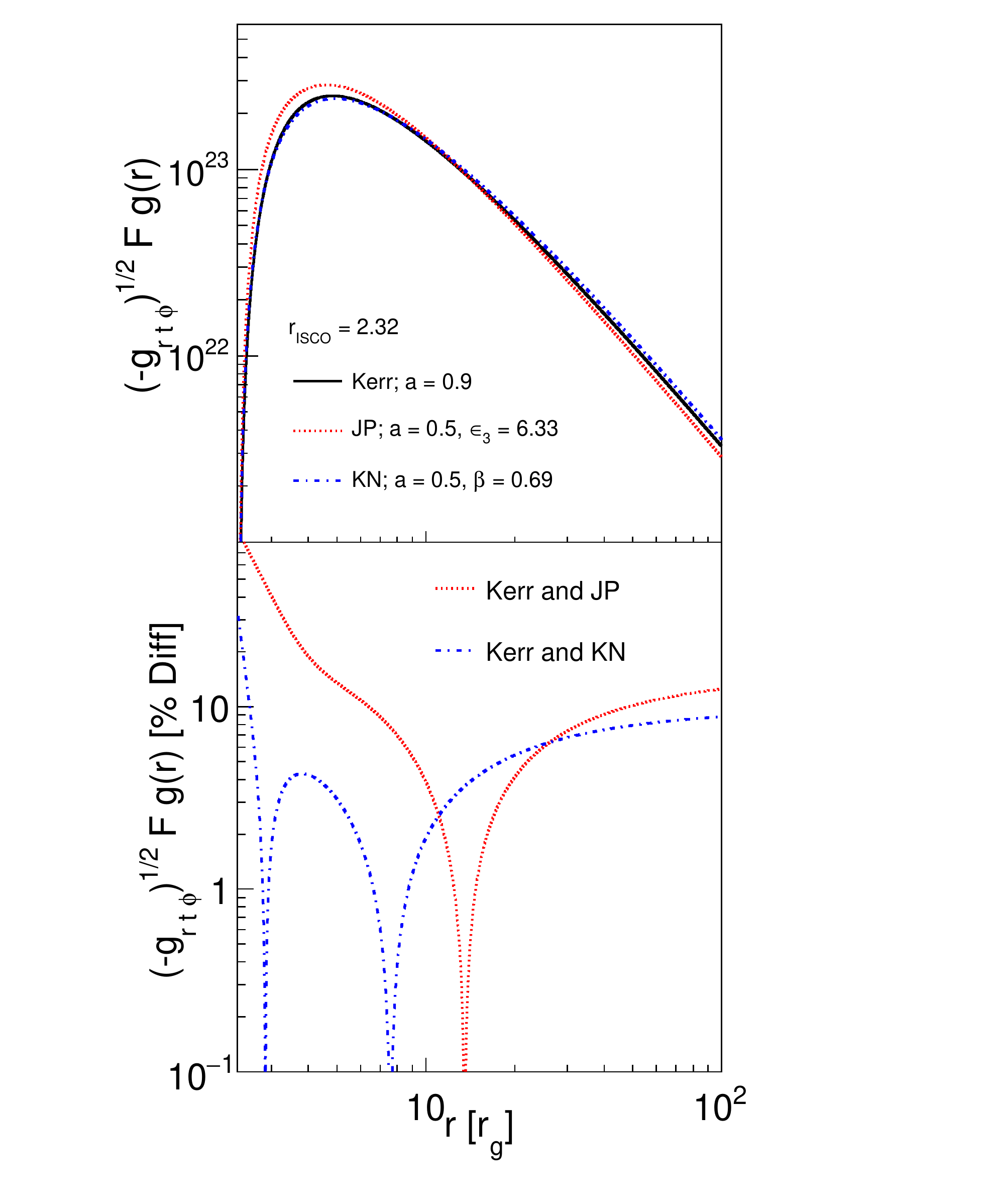}
                \caption{\label{fig:power}} 
        \end{subfigure}
        \caption{\label{fig:fluxPlots}Radial flux (eq \ref{eq:flux}) (\ref{fig:flux}) and power (\ref{fig:power}) for the JP, KN, and Kerr metrics which all give $r_{ISCO}=$2.32 $r_g$. The bottom panels show the comparison of the alternative metrics to the Kerr metric (color online).}
	\end{figure}
	
		\begin{figure}
        \centering
        \begin{subfigure}[b]{0.21\textwidth}
                \includegraphics[trim=35mm 0mm 30mm 0mm ,width=\textwidth]{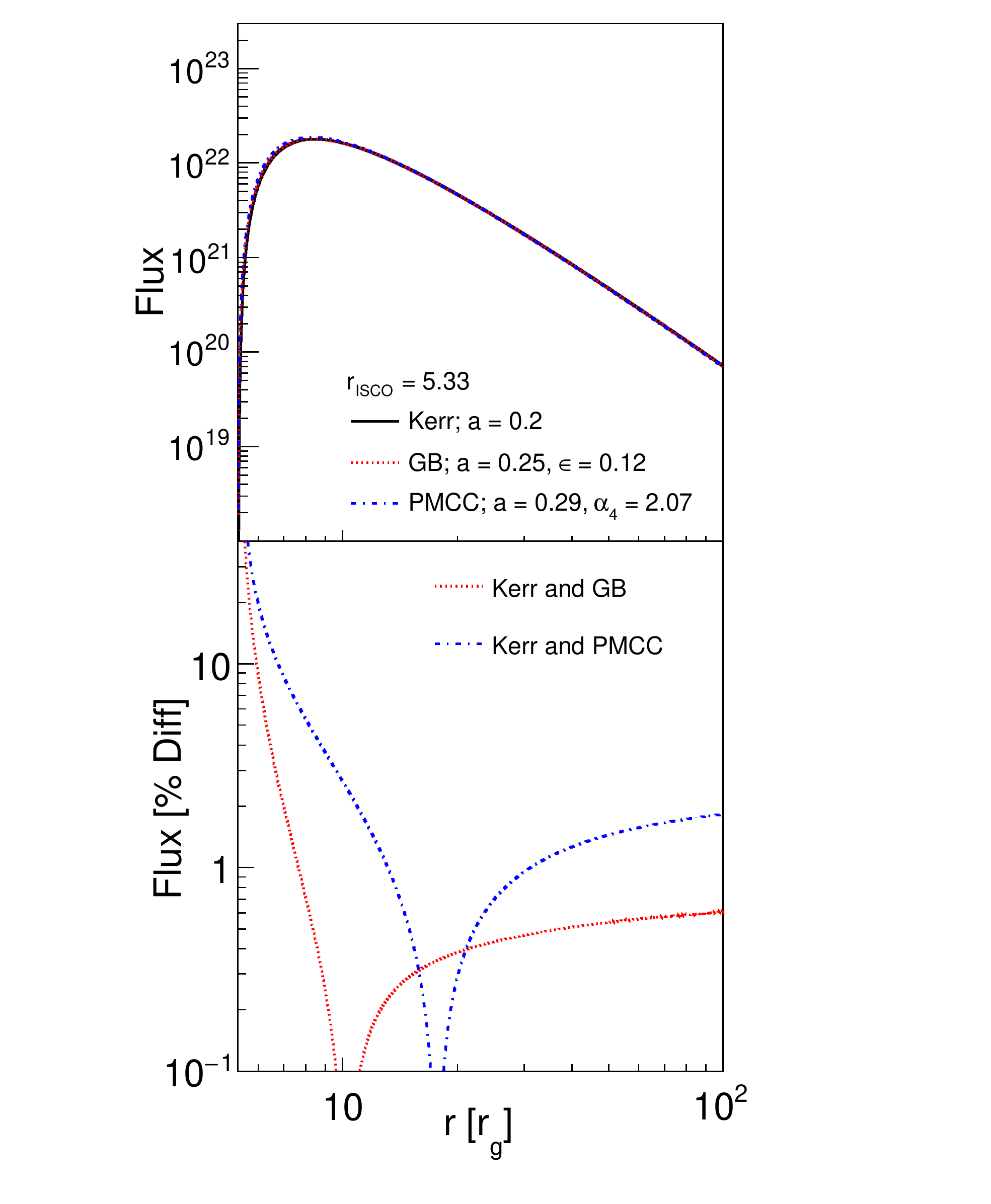}
                \caption{\label{fig:fluxLS}}
        \end{subfigure}
         \quad
        \begin{subfigure}[b]{0.21\textwidth}
                \includegraphics[trim=30mm 0mm 35mm 0mm ,width=\textwidth]{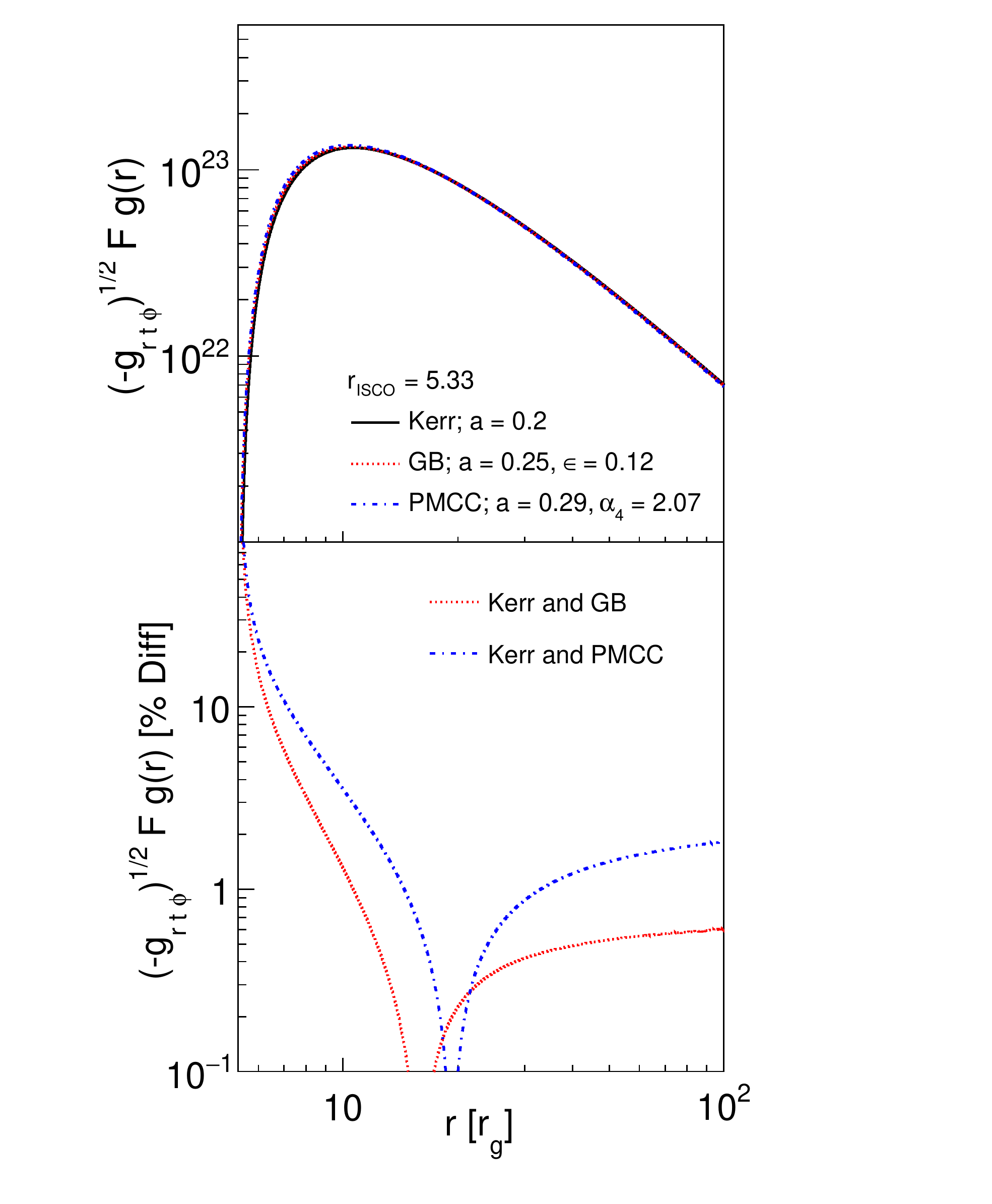}
                \caption{\label{fig:powerLS}} 
        \end{subfigure}
        \caption{\label{fig:fluxPlotsLS}Radial flux (eq \ref{eq:flux}) (\ref{fig:flux}) and power (\ref{fig:power}) for the GB, PMCC, and Kerr metrics which all give $r_{ISCO}=$5.33 $r_g$.  The bottom panels show the comparison of the alternative metrics to the Kerr metric (color online).}
	\end{figure} 
	
The left panels of Figs. \ref{fig:fluxPlots} and \ref{fig:fluxPlotsLS} compare the fluxes $F(r)$ emitted in the plasma frame for the different metrics. For the rapidly spinning black holes (Fig. \ref{fig:flux}), the fractional differences in $F(r)$ are typically a few percent.
The difference is larger for the innermost part of the accretion flow with the Kerr $F(r)$ exceeding the values of the non-Kerr metrics by up to 30\%.  

The right panels of the figures show the power $P$ emitted per unit Boyer Lindquist time and per Boyer Lindquist 
radial interval $dr$:   
\begin{equation}
\frac{dP}{dr}(r)\,=\,\sqrt{-g_{t r \phi}} F(r) g_{\rm em}^{\rm obs}
\end{equation}
The factor $\sqrt{-g_{t r \phi}}$ is the $t-r-\phi$ dependent part of the metric and is used to transform 
the {\it number} of emitted photons per plasma frame $d\hat{t}$ and $d\hat{r}$ into that emitted 
per Boyer Lindquist $dt$ and $dr$ \cite{Kulkarni2011}. 
The last factor corrects for the frequency change of the photons between their emission in the plasma rest frame and their detection by an observer at infinity. We estimate the effective redshift between emission and observation by assuming photons are emitted in the
upper hemisphere with the dimensionless wave vector 
$\hat{k}^{\mu}\,=\,(1,0,-1,0)$ in the plasma frame. After transforming $\hat{k}$ 
into the wave vector $k$ in the Boyer Lindquist frame we calculate the photon energy at infinity $E_{\gamma}$ from the 
constant of motion associated with the time translation Killing vector $(1,0,0,0)$:
\begin{equation}
E_{\gamma}\,=\,-k_t   
\end{equation}
and set $g_{\rm em}^{\rm obs}\,=\,E_{\gamma}$.
The different metrics exhibit very similar $dP/dr$-distributions with typical fractional differences of  $<$ 10\%. Again, the largest deviations are found near the ISCO.
Overall, the different metrics lead to very similar $F(r)$ and $dP/dr$-distributions because (i) we compare models
with identical $r_{\rm ISCO}$-values (leading radial profiles with a similar $r$-dependence), 
and (ii) we use fine-tuned accretion rates $\dot{M}$ to compensate for the different accretion 
efficiencies where $\eta = 1-E_{ISCO}$ (i.e. the different fractions of the rest mass energy that can be extracted when matter 
moves from infinity to $r_{\rm ISCO}$). In the following we focus on the rapidly spinning BH simulations, 
as the observables depend more strongly on the assumed background spacetime 
than for slowly spinning BHs.

	\begin{figure}
        \centering
        \begin{subfigure}[b]{0.22\textwidth}
                \includegraphics[trim=35mm 0mm 28mm 0mm, width=\textwidth]{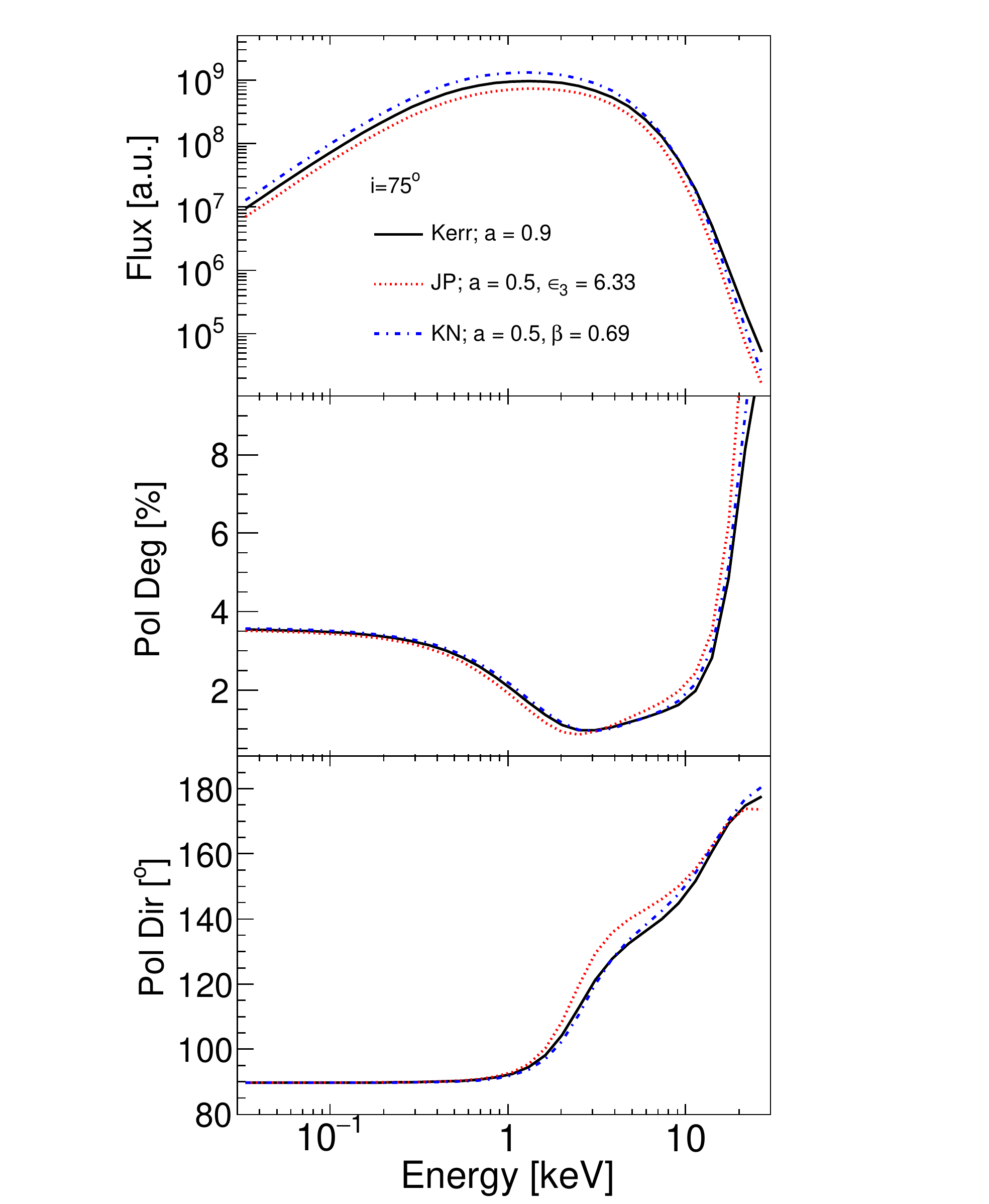}
                \caption{\label{fig:thermalHS}}
        \end{subfigure}
         \quad
        \begin{subfigure}[b]{0.22\textwidth}
                \includegraphics[trim=28mm 0mm 35mm 0mm, width=\textwidth]{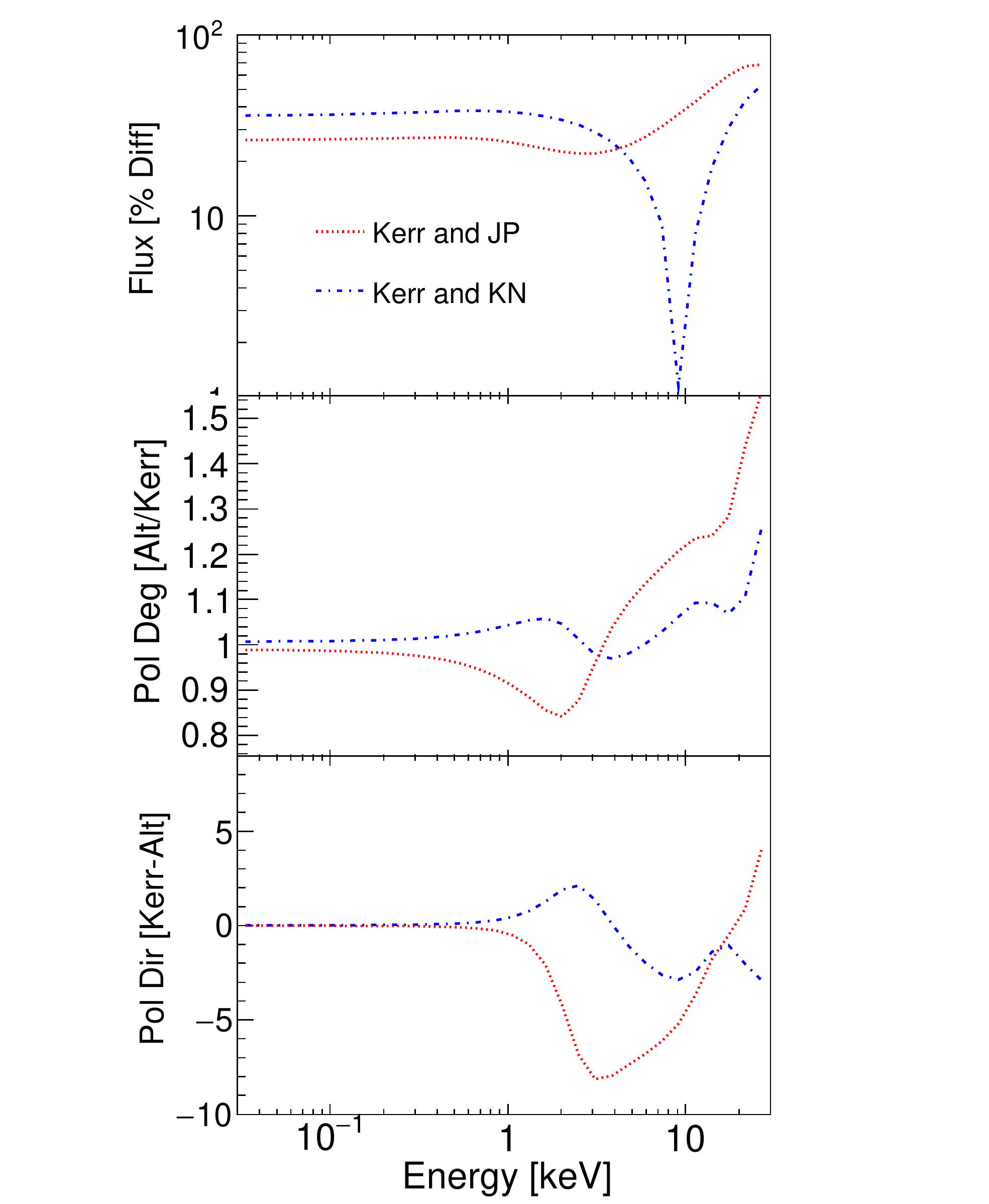}
                \caption{\label{fig:thermalHSPD}} 
        \end{subfigure}
       \caption{\label{fig:thermal}Flux (top panel), polarization fraction (middle panel), and polarization 	degree (bottom panel) of the thermal disk emission for the JP, KN, and Kerr metrics (\ref{fig:thermalHS}) and corresponding comparisons of the alternative metrics with respect to the Kerr metric (\ref{fig:thermalHSPD}) (color online).}
	\end{figure}
	
		\begin{figure}
        \centering
        \begin{subfigure}[b]{0.22\textwidth}
                \includegraphics[trim=35mm 0mm 28mm 0mm,width=\textwidth]{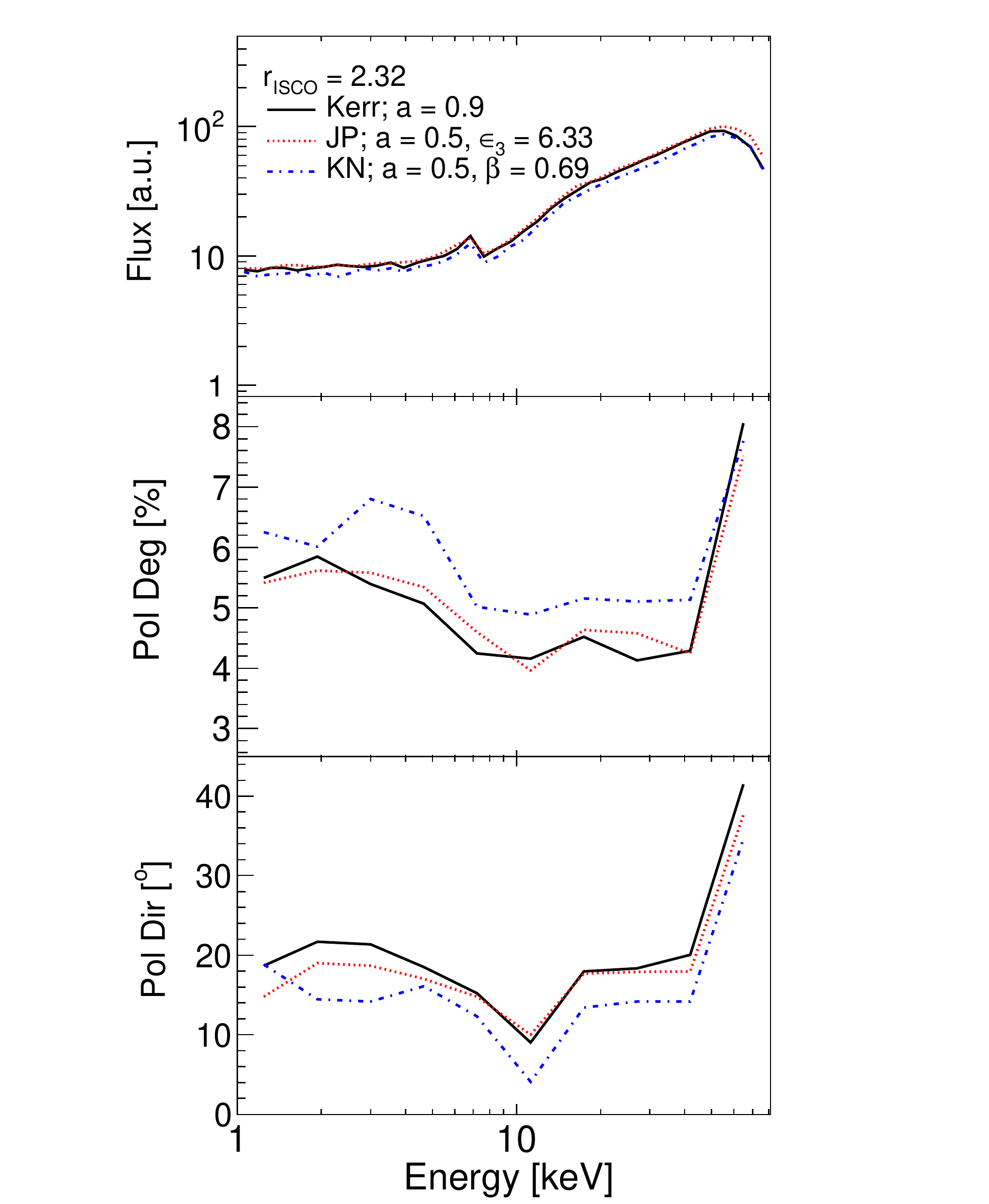}
                \caption{\label{fig:polI}}
        \end{subfigure}
         \quad
        \begin{subfigure}[b]{0.22\textwidth}
                \includegraphics[trim=28mm 0mm 35mm 0mm,width=\textwidth]{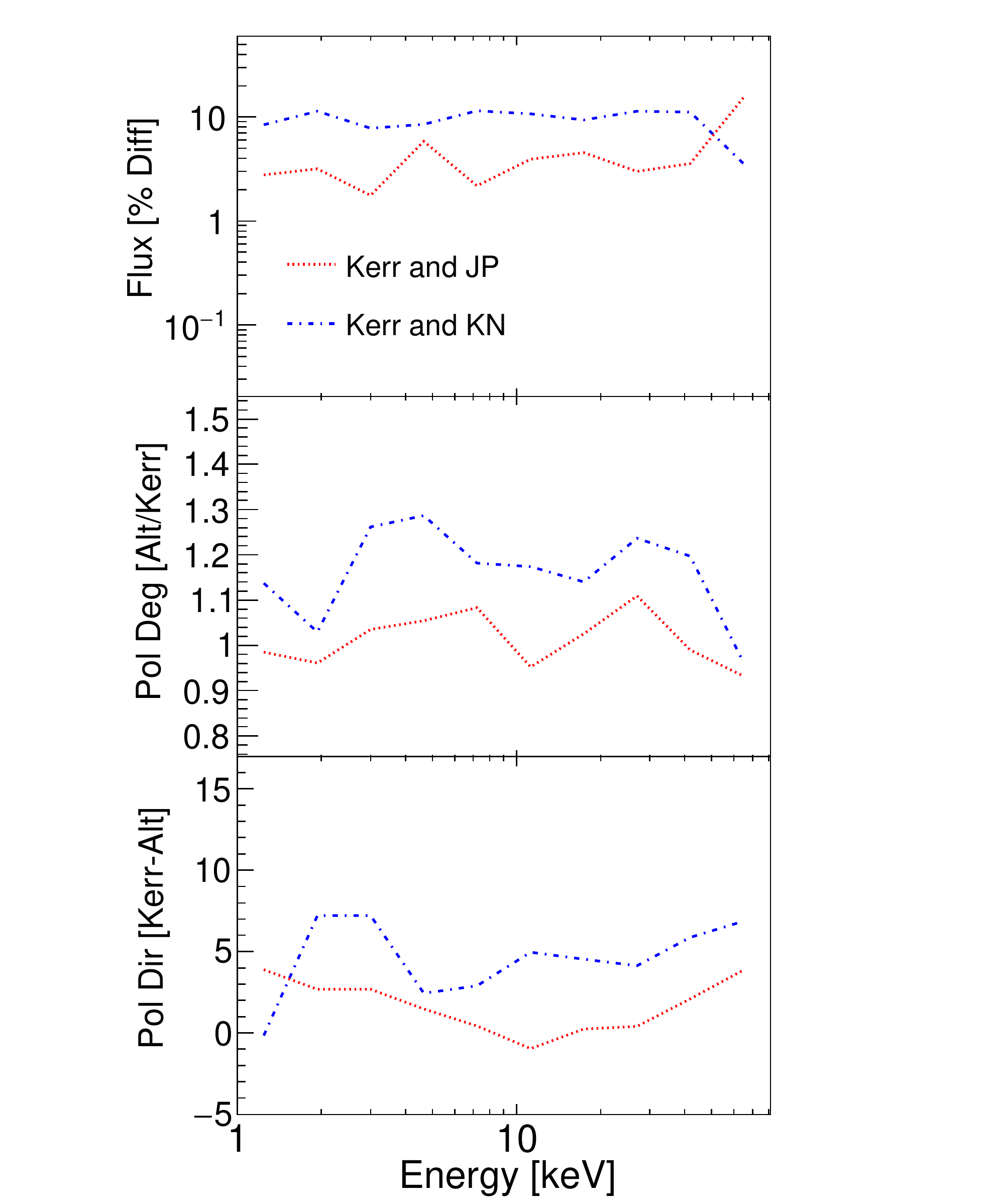}
                \caption{\label{fig:polIPD}} 
        \end{subfigure}
       \caption{\label{fig:plI}Flux (top panel), polarization fraction (middle panel), and polarization degree (bottom panel) for the reflected spectrum showing the Fe-K$\alpha$ line of an AGN with $h=3 r_g$ for the JP, KN, and Kerr metrics (\ref{fig:polI}) and corresponding comparisons of the alternative metrics with respect to the Kerr metric (\ref{fig:polIPD})(color online).}
	\end{figure}	

The analysis presented here focuses on specific choices for matching the Kerr metric with non-Kerr counterparts.
This matching is not unique as the non-Kerr metrics give the same $r_{\rm ISCO}$ for a continuous family of different
metrics. We simulated a few non-Kerr metrics giving the same $r_{\rm ISCO}$, and found that the differences between
these metrics and the Kerr metric are all comparable to the differences shown above.

Figure \ref{fig:thermal} shows the flux, polarization degree, and polarization direction of the thermal disk emission of a mass accreting stellar mass BH as shown for an observer at an inclination of 75$^{\circ}$.We only show the results for the rapidly spinning Kerr, JP and KN BHs.  While there are some differences in these spectra the overall shapes are similar.   At the highest energies, deep in the Wien tail of the multi-temperature energy spectrum, the fluxes show more differences owing to the different orbital velocities and thus Doppler boost of the emission and different fractions of photons reaching the observer versus photons falling into the BHs. The different metrics also lead to very similar polarization fractions and polarization angles.  However, in terms of the polarization properties the Kerr and KN metrics show almost identical results and the JP metric shows slightly different results. Overall, the main conclusion is that once we choose models with identical $r_{\rm ISCO}$ and correct for the different accretion efficiencies, the observational signatures depend only very  weakly on the considered metric.  Assuming that the background spacetime is described by the Kerr metric, 
the thermal energy spectrum and the polarization properties can be used to fit $r_{\rm ISCO}$ and the BH inclination $i$ \cite{Li2009, Schnittman2009}. The results presented so far
indicate that the fitted $r_{\rm ISCO}$ and $i$ values will not depend strongly on the assumed background spacetime. 

We now turn to the properties of the reflected corona emission from an AGN, assuming the lamp-post corona emits 
unpolarized emission with a photon power law index of $\Gamma = $ 1.7 from a height $h = $ 3 $r_g$ above the black hole 
(Fig. \ref{fig:plI}).  Again, the flux and polarization energy spectra are almost the same for all considered
metrics. The KN metric shows slightly larger deviations from the Kerr metric than the JP metric. 

Some accreting black holes exhibit quasi periodic oscillations (QPOs), i.e. peaks in the Fourier transformed power spectra.
The orbiting hot spot model \cite{Schnittman2004,Schnittman2005,Stella1998, Stella1999, Abramowicz2001, Abramowicz2003} explains the high-frequency QPOs (HFQPOs) 
of accreting stellar mass black holes with a hot spot orbiting the black hole close to the ISCO. 
If we succeeded to confirm the model (e.g. through the observations of the phase resolved energy spectra and/or polarization
properties \cite{Beheshtipour2015}), one could use HFQPO observations to measure the orbital periods close to the ISCO. 
In the case of AGNs, tentative evidence for periodicity associated with the ISCO has been found for several objects. 
Examples include orbital periodicity on the time scale of a few days as seen in the blazar OJ 287 \cite{Pihajoki2013} and QPO's at frequencies of O(100 Hz) such as that seen for the microquasar GRO J1655-40 \cite{Strohmayer2001}. 
\begin{figure}
        \begin{subfigure}[b]{0.22\textwidth}
                \includegraphics[trim=35mm 0mm 30mm 0mm ,width=\textwidth]{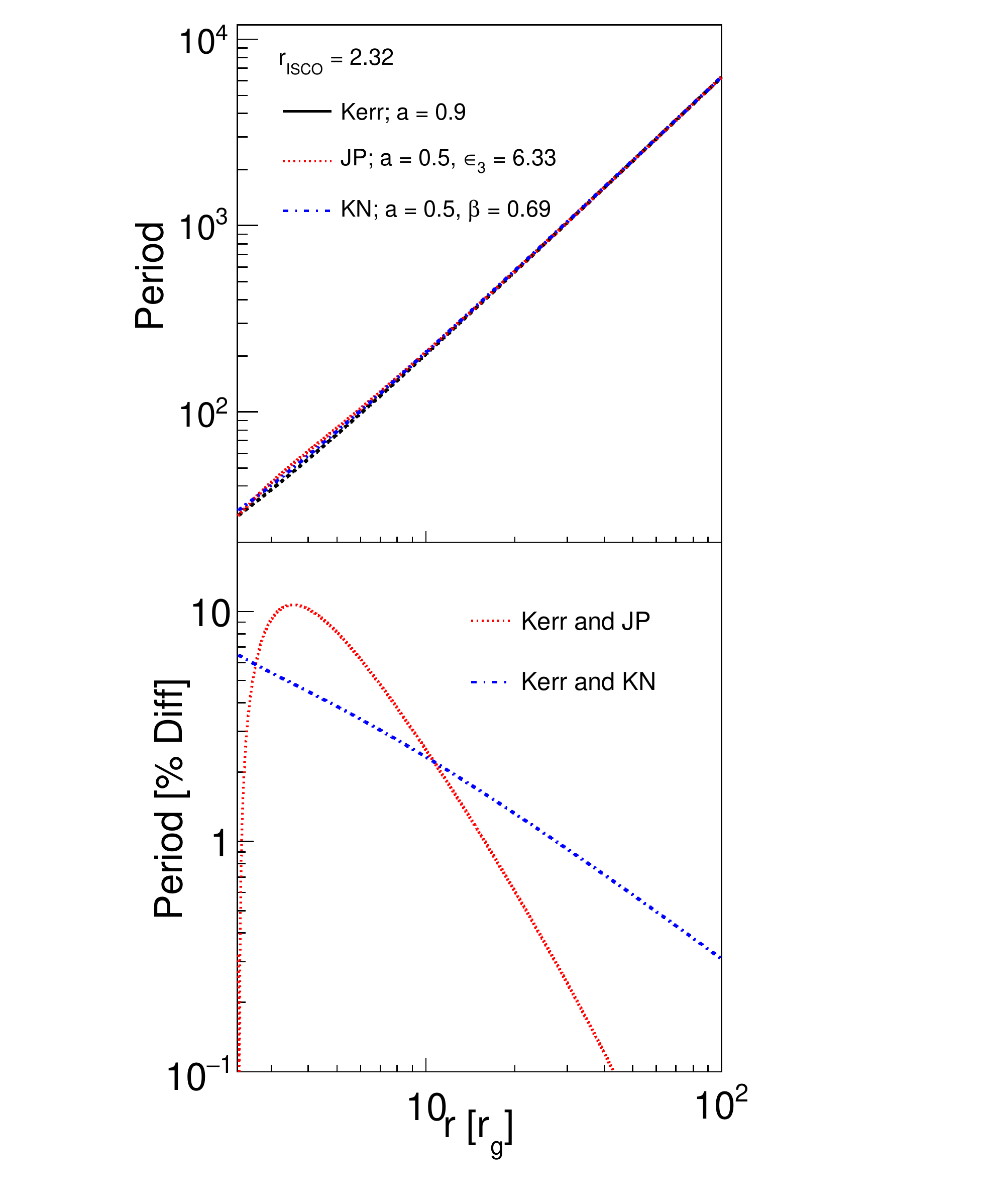}
                \caption{\label{fig:periodbig}}
        \end{subfigure}
         \quad
        \begin{subfigure}[b]{0.22\textwidth}
                \includegraphics[trim=30mm 0mm 35mm 0mm ,width=\textwidth]{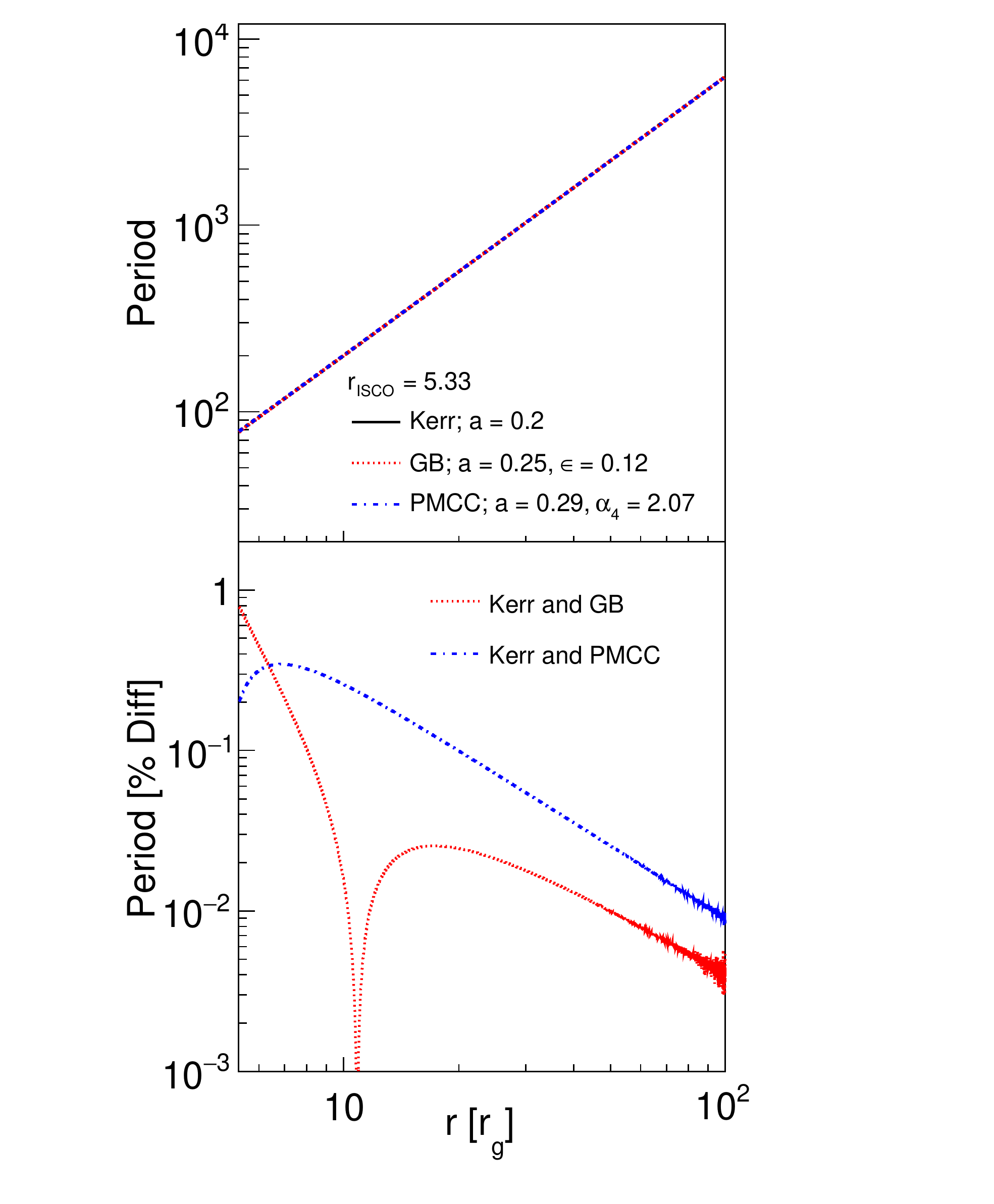}
                \caption{\label{fig:periodsmall}} 
        \end{subfigure}
        \caption{\label{fig:period} Orbital periods (top) for the metrics and the percent difference (bottom) for the JP and KN metrics compared to the Kerr metric (\ref{fig:periodbig}) and the GB and PMCC metrics (\ref{fig:periodsmall}) (color online)}
	\end{figure}
	\begin{figure} 
 		\includegraphics[trim=25mm 0mm 30mm 0mm ,width=.35\textwidth]{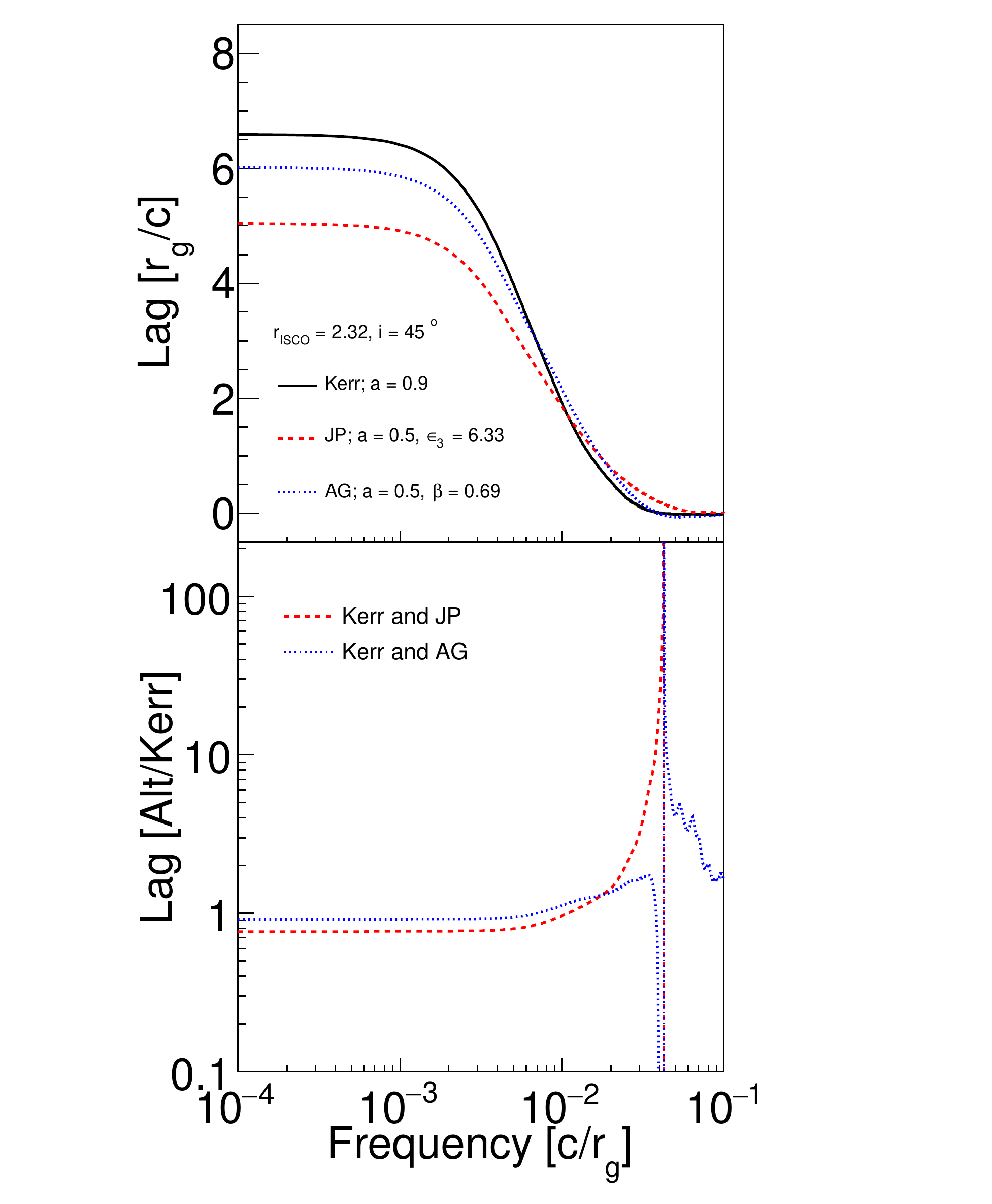} 
 		\caption{\label{fig:lag} Lag-frequency spectrum (top) of an AGN with $h=3 r_g$ for the JP, KN, and Kerr metrics along with the percent difference (bottom) of the KN and JP metrics when compared with the Kerr metric. A positive lag corresponds to the reflected emission lagging behind the direct emission where the direct emission is given in the 1-2 keV band and the reflected emission is in the 2-10 keV band (color online).}
	\end{figure}
Figure \ref{fig:period} shows that different metrics do predict different orbital periods which vary by up to $\sim$ 10\%. 

We investigated if other {\it timing properties} can be used to observationally distinguish between the different metrics by 
analyzing the observable time lags between the direct corona emission and the reflected emission assuming the lamp post geometry.
We use the standard X-ray reverberation analysis methods described by \cite{Uttley2014}. As expected, 
the 2-10 keV flux variations lag the 1-2 keV flux variations (Figure \ref{fig:lag}). 
	The JP metric leads to time lags up to 24\% shorter than the 
Kerr and KN metrics at low frequencies.  At frequencies between 0.01 $c/r_g$ and 0.1 $c/r_g$  
phase wrapping begins to occur (when the lag changes sign and begins to oscillate around 0) 
leading to the larger differences seen in this range.

\begin{figure*}
        \centering
        \begin{subfigure}[b]{0.4\textwidth}
                \includegraphics[trim=5mm 0mm 5mm 0mm, width=\textwidth]{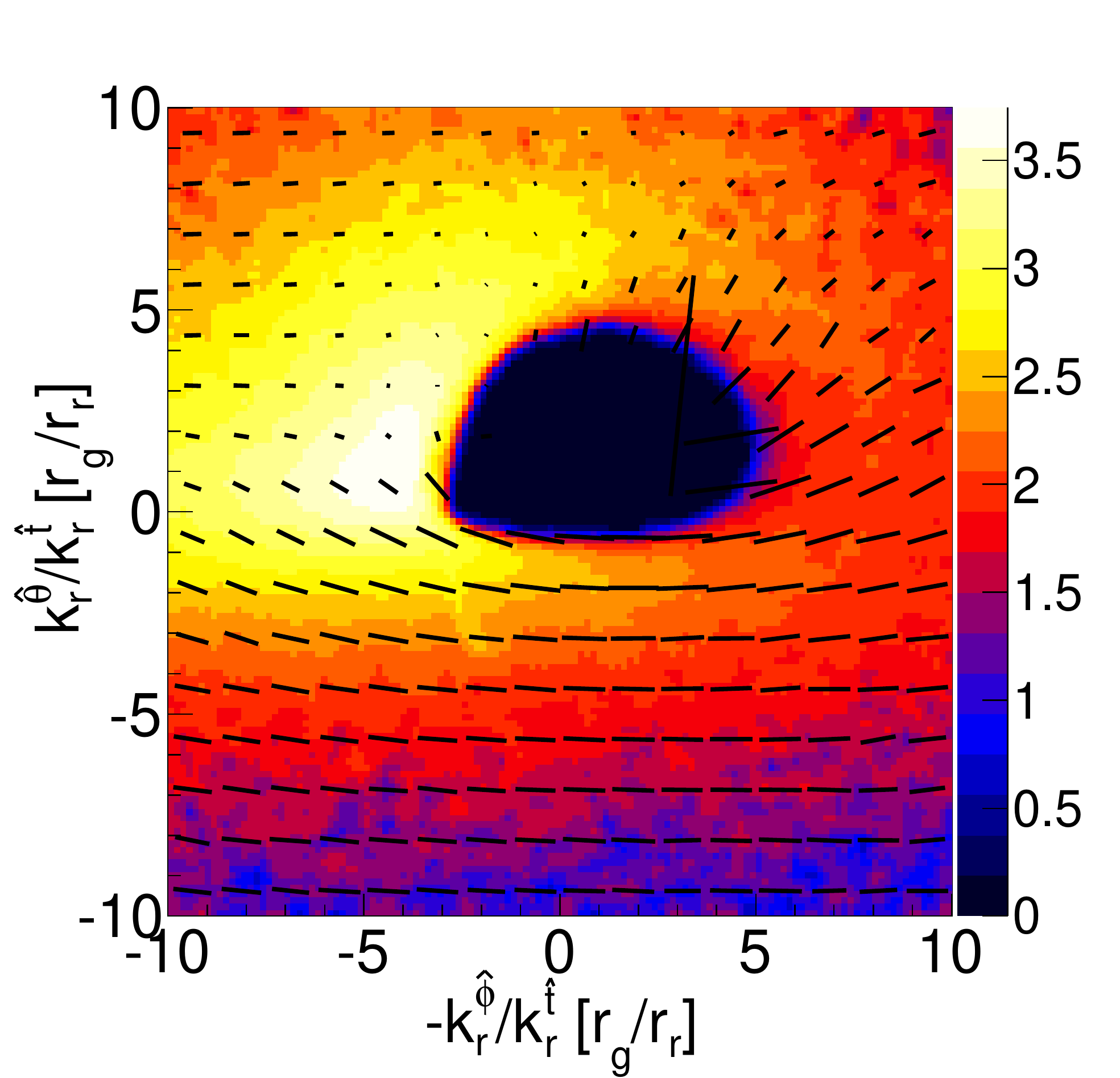}
        \end{subfigure}
         \quad
        \begin{subfigure}[b]{0.4\textwidth}
                \includegraphics[trim=5mm 0mm 5mm 0mm ,width=\textwidth]{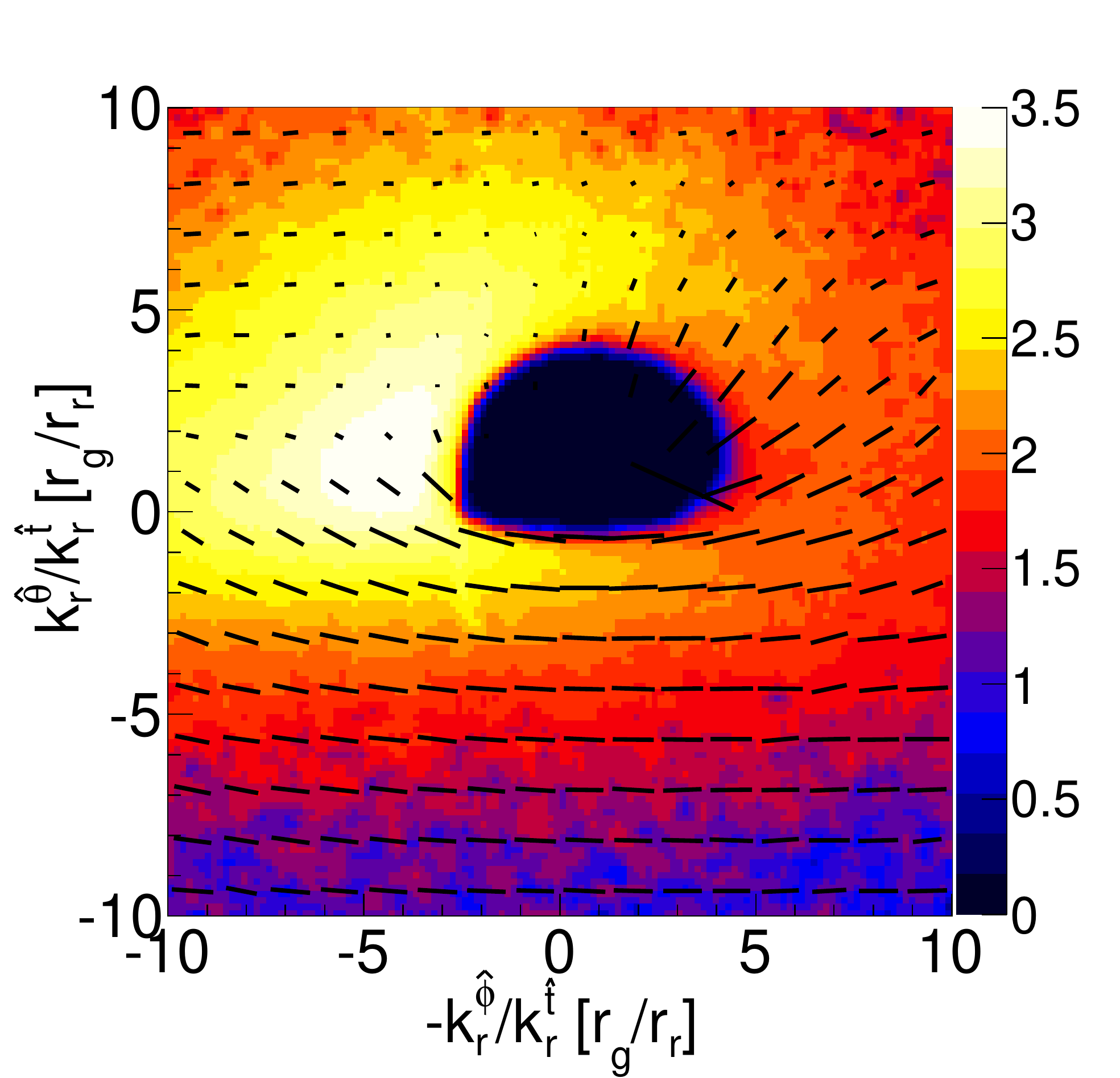}
        \end{subfigure}
        
        \begin{subfigure}[b]{0.4\textwidth}
                \includegraphics[trim=5mm 0mm 5mm 0mm ,width=\textwidth]{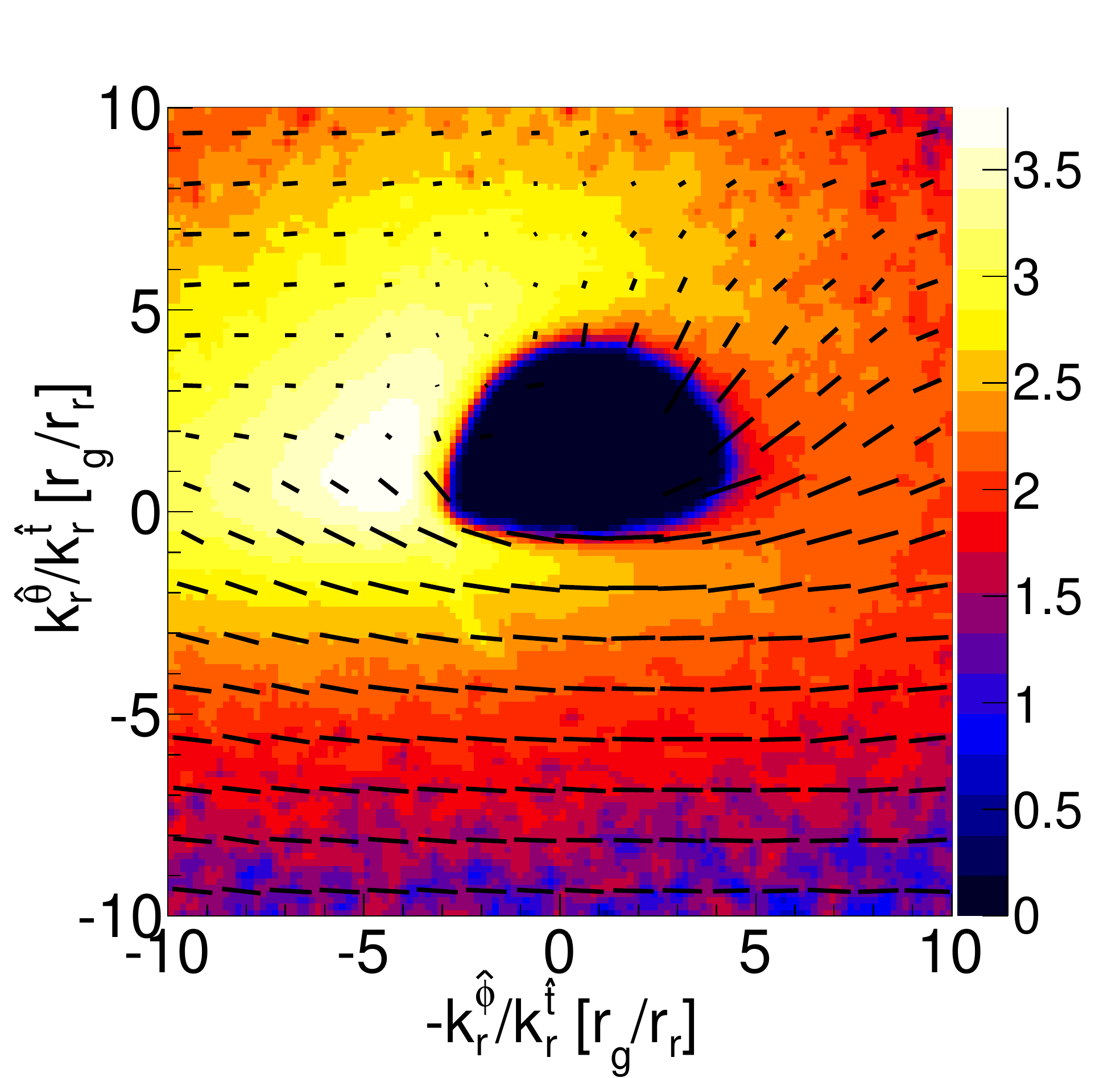}
        \end{subfigure}
        \quad
        \begin{subfigure}[b]{0.4\textwidth}
                \includegraphics[trim=5mm 0mm 5mm 5mm,width=\textwidth]{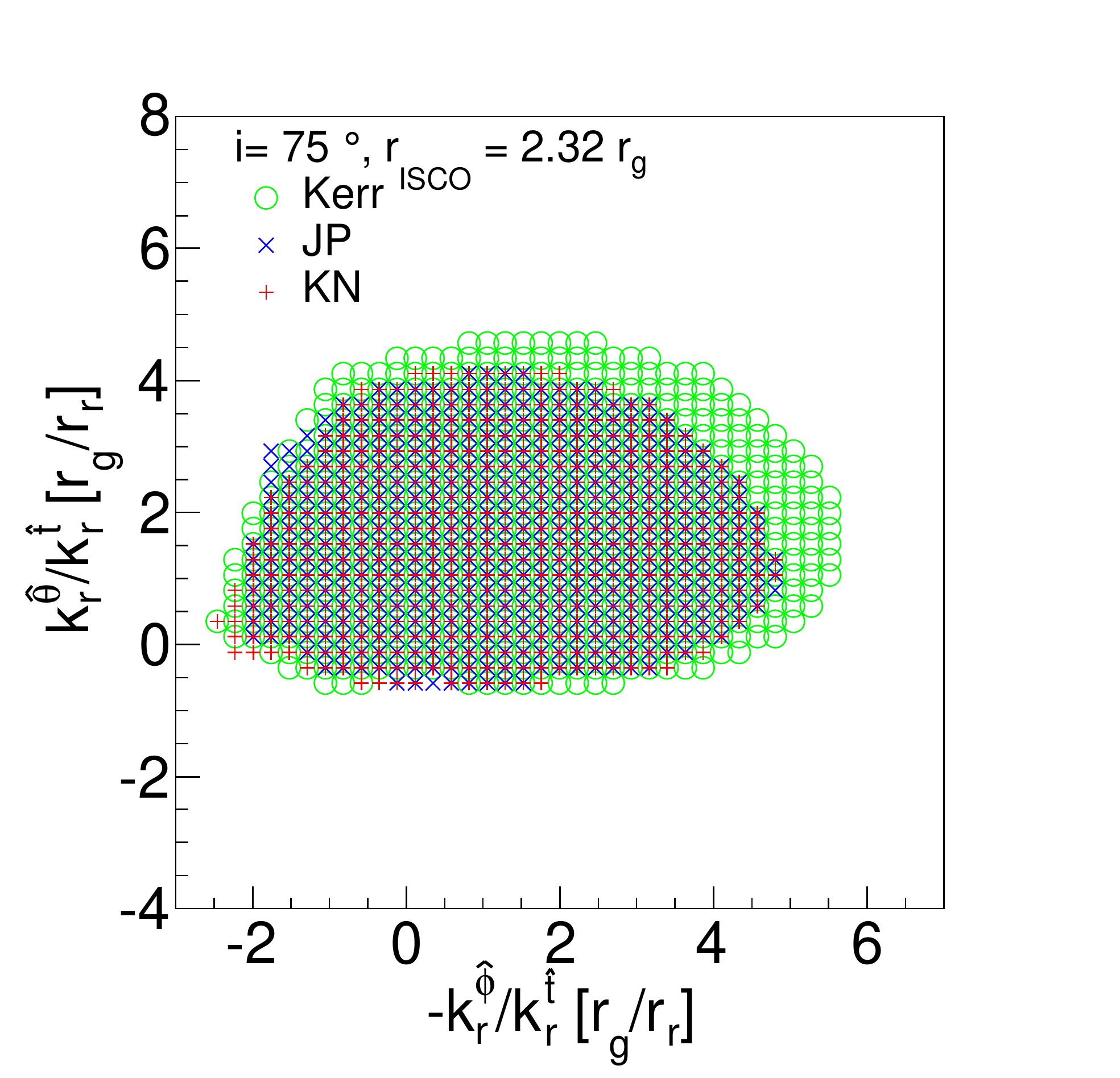}
        \end{subfigure}
        \caption{\label{fig:set1hi}Intensity-Polarization maps of the three rapidly spinning metrics in: Kerr (top left), JP (top right), KN (bottom left) for a black hole at a high inclination along with a comparison of their shadows (bottom right) (color online.)}
	\end{figure*}

\begin{figure}
        \centering
        \begin{subfigure}[b]{0.35\textwidth}
                \includegraphics[trim=15mm 0mm 15mm 0mm ,width=\textwidth]{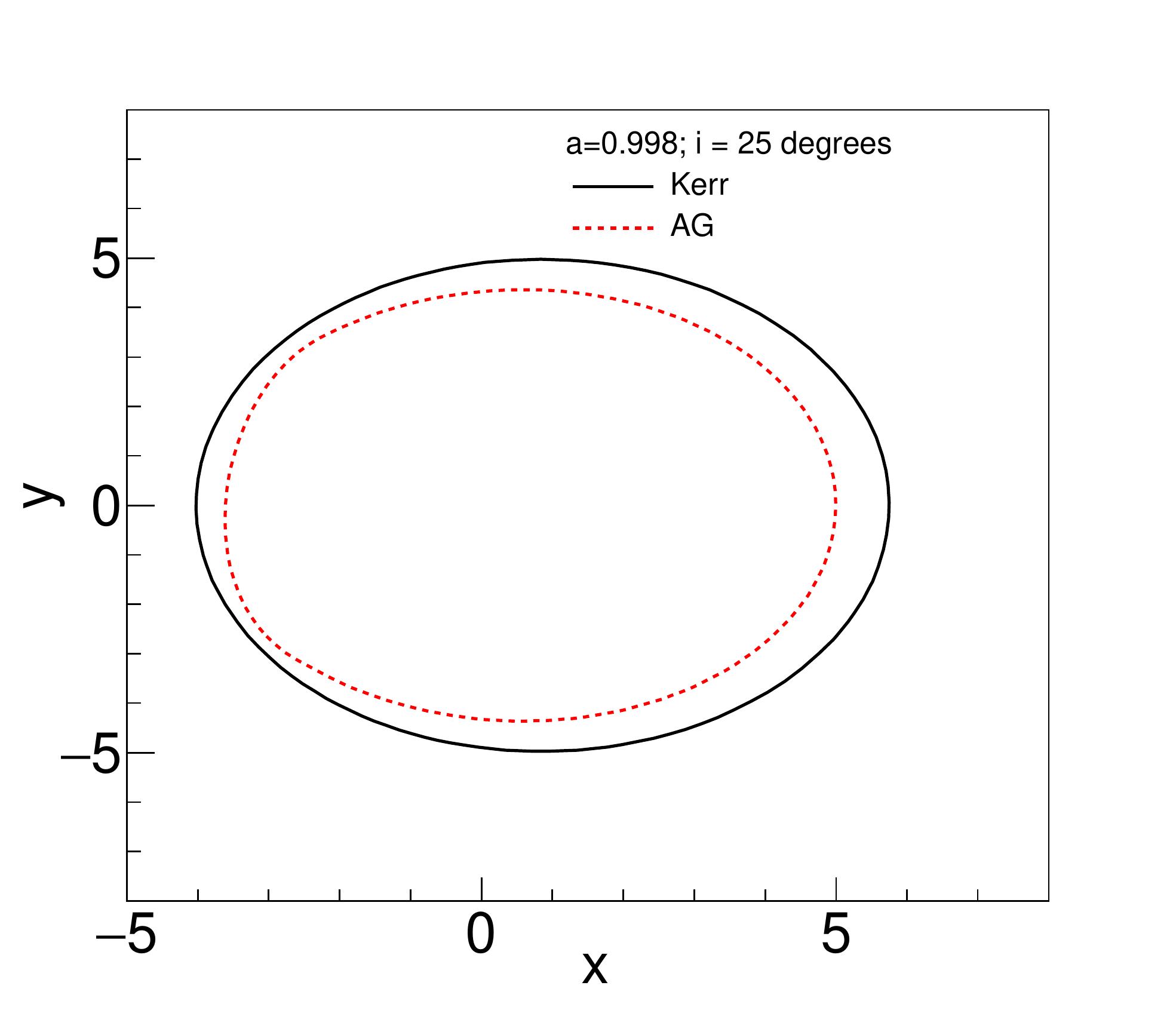}
                \caption{\label{fig:ring25}}
        \end{subfigure}
         \quad
        \begin{subfigure}[b]{0.35\textwidth}
                \includegraphics[trim=15mm 0mm 15mm 0mm ,width=\textwidth]{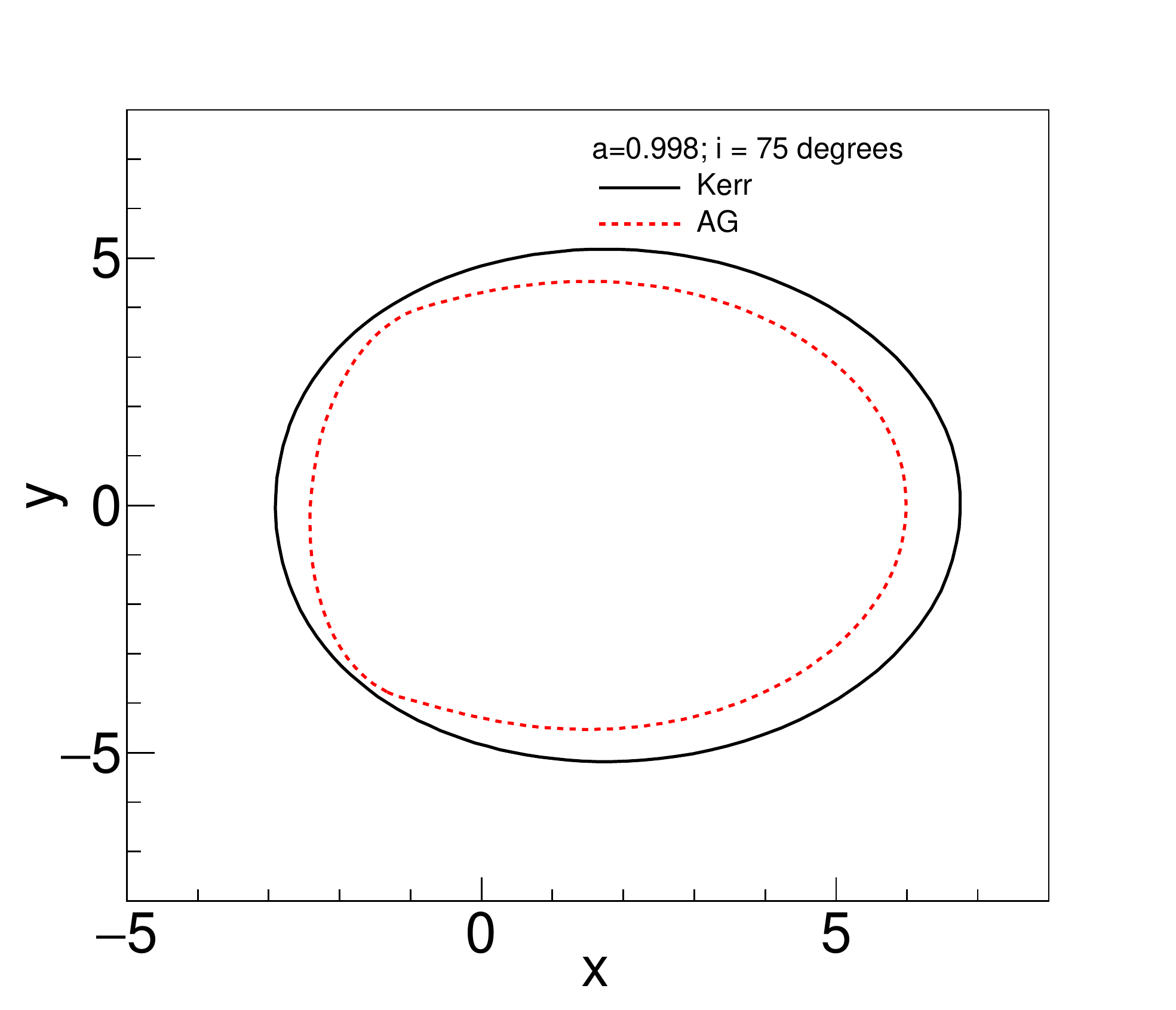}
                \caption{\label{fig:ring75}} 
        \end{subfigure}
       \caption{\label{fig:rings}Photon rings of the Kerr and KN metrics as seen by an observer at both $i=25 ^{o}$ (\ref{fig:ring25}) and $i=75^{o}$ (\ref{fig:ring75}) (color online).}
	\end{figure}

Although the considered metrics give the same $r_{\rm ISCO}$ in Boyer-Lindquist coordinates, the black hole 
shadow may have a different shape and/or size when viewed by an observer at infinity \cite[see][for a related study]{Johannsen2010,Johannsen2013c}. The results for the Kerr, JP, and KN metrics are shown in Fig. \ref{fig:set1hi} 
for an inclination of $i=75^{\circ}$.  The shadow of the KN and JP metrics is $\sim$15\% smaller than that 
of the Kerr metric. Furthermore, the shapes differ slightly. Similarly, the shapes of the photon rings are shown in Fig. \ref{fig:rings} which are calculated by following the procedures outlined in Section III of \cite{Bardeen1973}.

\section{Summary and Discussion}

In this paper we studied the observational differences between accreting black holes in five different background spacetimes, including GR's Kerr spacetime, and four alternative spacetimes. We chose the  parameters of the considered metrics 
as to give identical innermost stable orbits in Boyer Lindquist coordinates. 
The predicted observational differences are larger for rapidly spinning black holes. 
Overall the observational differences are very small if we adjust the accretion rate to correct for the metric-dependent 
accretion efficiency. 
The measurement of the predicted differences are very small -- especially if one accounts for the astrophysical uncertainties, 
i.e. observational and theoretical uncertainties of the accretion disk properties. 
From an academic standpoint, it is interesting to compare the small differences of the predicted properties, e.g. the 
differences of the thermal energy spectra and the BH shadow images.  The thermal spectrum of the Kerr BH is slightly 
harder than that of the JP and KN black holes (for the same $r_{\rm ISCO}$),
and the Kerr BH shadow is slightly larger than that the JP and KN BH shadows.
Reducing the spin of the Kerr BH would make the spectral difference smaller, but would increase the mismatch between 
the apparent BH shadow diameters. Thus, in the absence of astrophysical uncertainties, the combined information from various observational channels could be used to distinguish between different metrics.

Although the analysis shows that the differences between the Kerr and non-Kerr metrics are rather subtle (especially
in the presence of uncertainties of the structure of astrophysical accretion disks), we can use existing observations 
to constrain large parts of the parameter space of the non-Kerr metrics (see also \cite{Bambi2014}).  As an example we use the recent observations 
of the accreting stellar mass black hole Cyg X-1 \cite{Gou2011}. The observations give a 3\,$\sigma$ 
upper limit of $r_{\rm ISCO}<1.94$. Inspecting Fig.\ref{fig:family} we see that the constraints on the ISCO can only
be fulfilled for JP deviation parameters $\epsilon_3 \in\left[-0.15 ,3.72\right]$ and KN deviation parameters 
$\beta\in \left[-0.01,0.73\right]$, excluding all parameter values outside of these intervals.
Our limits rest on the matching of Kerr to non-Kerr metrics via using identical $r_{\rm ISCO}$-values.  Maximally rotating black holes, when a $=$ 0.998 \cite{Thorne1974} provide the best opportunity to test GR because  observations of these systems can, in principle, be used to exclude all deviations from GR down to a small interval around 0, limited only by the actual spin value, the statistics of the observations, and the astrophysical uncertainties. 
A follow-up analysis could use state-of-the-art modeling of the actual X-ray data with accretion disk and
emission models in the Kerr and non-Kerr background spacetimes.

Further progress will be achieved by continuing to refine our understanding of BH accretion disks based on
General Relativistic Magnetohydrodynamic and General Relativistic Radiation Magnetohydrodynamic 
simulations \cite[e.g.][]{Schnittman2013a, Schnittman2013b, Sadowski2015} and matching simulated
observations to X-ray spectroscopic, X-ray polarization, and X-ray reverberation observations.   

The images of the BH shadow of Sgr A$^*$ with the Event Horizon Telescope can give additional constraints.
Whereas the images of the BH shadow still depend on the astrophysics of the accretion disk, 
imaging of the photon ring would be free of such uncertainties. Of course, much better imaging 
would be required to do so.

\begin{acknowledgments}
We would like to thank NASA (grant \#NNX14AD19G) and the Washington University McDonnell Center for the Space Sciences for financial support.
\end{acknowledgments}

\end{document}